%% file: GreedyCSC.tex
\documentclass{siamonline1116}

\input{rev_ex_shared}

\ifpdf
\hypersetup{
  pdftitle={\TheTitle},
  pdfauthor={\TheAuthors}
}
\fi

\begin{document}

\maketitle

\begin{abstract}
Sparse coding techniques for image processing traditionally rely on a processing of small overlapping patches separately followed by averaging. This has the disadvantage that the reconstructed image no longer obeys the sparsity prior used in the processing. For this purpose convolutional sparse coding has been introduced, where a shift-invariant dictionary is used and the sparsity of the recovered image is maintained. Most such strategies target the $\ell_0$ ``norm'' or the $\ell_1$ norm of the whole image, which may create an imbalanced sparsity across various regions in the image. In order to face this challenge, the $\ell_{0,\infty}$ ``norm'' has been proposed as an alternative that ``operates locally while thinking globally". The approaches taken for tackling the non-convexity of these optimization problems have been either using a convex relaxation or local pursuit algorithms. In this paper, we present an efficient greedy method for sparse coding and dictionary learning, which is specifically tailored to $\ell_{0,\infty}$, and is based on matching pursuit. We demonstrate the usage of our approach in salt-and-pepper noise removal and image inpainting.
\end{abstract}

\begin{keywords}
Convolutional Sparse Coding, Global Modeling, Local Processing, Greedy Algorithms, Sparse Representations
\end{keywords}

\begin{AMS}
  68U10, 62H35, 94A08
\end{AMS}

\section{Introduction}
Sparse coding can be described as solving the following minimization problem, known as the $P_0$ problem \cite{P0}:
\[
\left( P_0 \right):~~~\underset{\boldsymbol{\alpha}}{\text{min}}  \left\Vert \boldsymbol{\alpha} \right\Vert_0 ~s.t.~\textbf{x}=\text{D}\boldsymbol{\alpha},
\]
where $\boldsymbol{\alpha} \in \mathbb{R}^p$ is a sparse representation of a signal $\textbf{x} \in \mathbb{R}^N$ in the dictionary $\text{D} \in \mathbb{R}^{N \times p}$. The columns of $\text{D}$, which are referred to as atoms, are a full and overcomplete set, and we will assume without loss of generality that they are normalized to unit $\ell_2$ norm. The $\ell_0$ ``norm''\footnote{The $\ell_0$ ``norm'' is not actually a norm, but we shall nonetheless refer to it as a norm for the sake of brevity.} returns the number of nonzero elements in a vector, also called the sparsity.

When modeling natural images, we allow some deviation $\epsilon$ from the model rather than seeking a perfect reconstruction:
\[
\left( P_0^\epsilon \right):~~~\underset{\boldsymbol{\alpha}}{\text{min}} \left\Vert \boldsymbol{\alpha} \right\Vert_0 ~s.t.~\left\Vert \textbf{x} - \text{D}\boldsymbol{\alpha} \right\Vert_2^2 \leq \epsilon.
\]
An alternative form, in which the sparsity $k$ is known, is:
\[
\left( P_0^k \right):~~~\underset{\boldsymbol{\alpha}}{\text{min}} 
\left\Vert \textbf{x} - \text{D}\boldsymbol{\alpha} \right\Vert_2^2 ~s.t.~ \left\Vert \boldsymbol{\alpha} \right\Vert_0 \leq k.
\]

As the $P_0^\epsilon$ and $P_0^k$ problems are NP-hard, several approximation techniques have been proposed. Matching Pursuit (MP) \cite{MP} is a greedy algorithm that in each iteration updates the representation with the coefficient that decreases the squared representation error the most. This coefficient is found by computing the inner products between the signal and all atoms in the dictionary $\text{D}$, selecting the one with the largest absolute value, and adding the inner product to the corresponding element in the sparse representation vector. Orthogonal Matching Pursuit (OMP) \cite{OMP} is a similar method, in which after each iteration, the set of all nonzero coefficients selected so far are updated to minimize the squared representation error. 
Thresholding (\cite[Chapter 4]{MBook}) is a very simple greedy algorithm, which computes the inner products between the signal and the atoms only once, and selects the nonzero elements in the representation as those corresponding to their largest absolute values.

Another class of algorithms uses a convex relaxation of the $\ell_0$ ``norm'' to the $\ell_1$ norm, known as the $P_1$ problem \cite{P0}. Many algorithms have been proposed for solving its unconstrained form (e.g. \cite{FISTA}, \cite{ISTA}, \cite{P1}), which reads as
\[
\left( P_1^\lambda \right):~~~\underset{\boldsymbol{\alpha}}{\text{min}} 
\left\Vert \textbf{x} - \text{D}\boldsymbol{\alpha} \right\Vert_2^2 + \lambda \left\Vert \boldsymbol{\alpha} \right\Vert_1 
\]
and is also known as LASSO \cite{LASSO} or Basis Pursuit Denoising (BPDN) \cite{BPDN}.

A useful measure for analyzing the behavior of the sparse coding problem is the mutual coherence of the dictionary, which measures the similarity between the atoms. It is defined as:
\[
\mu\left( \text{D} \right) = \underset{1 \leq i,j \leq p, i \neq j}{\text{max}} \left\vert \textbf{d}_i^T \textbf{d}_j \right\vert.
\]
In the noiseless case, the minimizer of the $P_1$ problem has been shown to coincide with the minimizer of the $P_0$ problem for sufficiently sparse representations \cite{Donoho}. If a solution to $\textbf{x}=\text{D}\boldsymbol{\alpha}$ exists and obeys
\[
\left\Vert \boldsymbol{\alpha} \right\Vert_0 < \frac{1}{2} \left( 1 + \frac{1}{\mu\left( \text{D} \right)}\right),
\]
then it is the unique solution of both $P_0$ and $P_1$. In such a case, OMP is also guaranteed to recover it exactly.

The dictionary $\text{D}$ may be an analytically defined matrix or operator. Yet, learning it from examples may provide sparser solutions and thus, better performance in various applications \cite{Rubinstein}. When training a dictionary on a set of vectors $\left\{ \textbf{x}_i \right\}_{i=1}^{s}$, one typically solves the minimization problem
\begin{equation}
\underset{\text{D},\left\{ \boldsymbol{\alpha}_i \right\}_{i=1}^{s}}{\text{min}}~~\overset{s}{\underset{i=1}{\sum}} \left\Vert \textbf{x}_i - \text{D}\boldsymbol{\alpha}_i \right\Vert_2^2  ~s.t.~\left\Vert \boldsymbol{\alpha}_i \right\Vert_0 \leq k~,~1 \leq i \leq s,
\label{eq:DL1}
\end{equation}
or, alternatively, the problem of minimizing the sparsity with a constraint on the representation error:
\begin{equation}
\underset{\text{D},\left\{ \boldsymbol{\alpha}_i \right\}_{i=1}^{s}}{\text{min}}~~\overset{s}{\underset{i=1}{\sum}} \left\Vert \boldsymbol{\alpha}_i \right\Vert_0  ~s.t.~\left\Vert \textbf{x}_i - \text{D}\boldsymbol{\alpha}_i \right\Vert_2^2 \leq \epsilon~,~1 \leq i \leq s.
\label{eq:DL2}
\end{equation}

Notice that if we relax the $\ell_0$ ``norm'' to the convex $\ell_1$ norm, the problems  (\ref{eq:DL1}) and (\ref{eq:DL2}) become convex only in $\text{D}$ and $\left\{ \boldsymbol{\alpha}_i \right\}_{i=1}^{s}$ separately, but not jointly. A common approach to solve the dictionary learning problem is by repeatedly alternating between optimizing the dictionary with the sparse representations held fixed, and optimizing the sparse representations with the dictionary held fixed. Optimizing the sparse representation vectors can be done to each signal separately using either $\ell_1$ based methods or greedy methods such as MP. Several dictionary learning methods have been proposed, including MOD \cite{MOD} and K-SVD \cite{KSVD}. 

In practice, these dictionary learning methods are only learned on relatively small signals due to computational complexity, memory requirement and the required quantity of training signals. When each $\textbf{x}_i$ is a $\sqrt{N} \times \sqrt{N}$ image, the computational cost of the sparse coding step for each image using MP is $O\left( Npk \right)$, and the computational cost of the K-SVD dictionary update step is $O\left( Np^2+kN+pk \right)$ \cite{KSVDcost}. Since $p$ is of the order of $N$, the dictionary learning is not scalable for full images. The current common practice is to divide the image into small patches of size $\sqrt{n} \times \sqrt{n}$, where each patch has its own sparse representation in a local dictionary $\text{D}_L \in \mathbb{R}^{n \times p}$ and now $p$ is of the order of $n$ (patch size) and not $N$ (image size). The computational complexity of the sparse coding step for overlapping patches where each patch has a sparsity of $k_p$ is $O\left( Nnpk_p \right)$, and the complexity of the K-SVD dictionary update step is $O\left( np^2+k_pNmn+Npk_p \right)$.
While some solutions have been proposed for working with large dimensions \cite{Alexandroni}, \cite{Double}, \cite{TruncatedWavelet}, they mainly allow working with larger patches or with sampling that is not on the grid.

\subsection{Convolutional sparse coding}
An appealing model that addresses these problems has been introduced in the form of convolutional sparse coding \cite{fast}, \cite{fastproximal}, \cite{shiftinvariant}, \cite{Heide}, \cite{Moreau}, \cite{deconv}, which has demonstrated impressive performance in various applications \cite{super}, \cite{fusion}, \cite{Osendorfer}, \cite{compressed}, \cite{ZhangPatel}. With this strategy, a set of local atoms $\left\{ \textbf{d}_j \right\}_{j=1}^p$ is used to represent a global signal by convolutions with representation vectors. Assuming that the representation in these dictionaries is sparse, we get the following optimization problem:
\begin{equation}
\underset{\left\{ \boldsymbol{\alpha}_j \right\}}{\text{min}} ~ \left\Vert \textbf{x} - \overset{p}{\underset{j=1}{\sum}} \textbf{d}_j \ast \boldsymbol{\alpha}_j \right\Vert_2^2 ~~s.t.~~  \overset{p}{\underset{j=1}{\sum}} \left\Vert \boldsymbol{\alpha}_j \right\Vert_0 \leq k.
\label{eq:CSC}
\end{equation}
Notice that we may recast this problem to be similar to $P_0$ by setting $\text{D}$ as the concatenation of Toeplitz matrices $\text{D}_j \in \mathbb{R}^{n \times p}$, $1 \leq j \leq p$, each representing a convolution with a kernel $\textbf{d}_j$:
\[ \textbf{x}=\text{D}\boldsymbol{\alpha}=\left[\begin{array}{cccc}\text{D}_1 & \text{D}_2 & ... & \text{D}_{p}
\end{array}\right]\left[\begin{array}{c}\boldsymbol{\alpha}_{1}\\\boldsymbol{\alpha}_{2}\\...\\\boldsymbol{\alpha}_{p}\end{array}\right] = \overset{p}{\underset{j=1}{\sum}} \text{D}_j \boldsymbol{\alpha}_j = \overset{p}{\underset{j=1}{\sum}} \textbf{d}_j \ast \boldsymbol{\alpha}_j, \]
where $\boldsymbol{\alpha}$ is the global sparse representation vector, $\text{D}$ is the global dictionary (also referred to as the convolutional dictionary), and $\boldsymbol{\alpha}_j$ is the vector of the coefficients multiplying $\textbf{d}_j$ at each of its shifts within the global signal. Each matrix $\text{D}_j$ is Toeplitz
for one-dimensional signals and is block-Toeplitz in the two-dimensional case (e.g., images). It is related to the signature dictionary from \cite{signature}, which is approximately shift invariant.

Greedy solutions to (\ref{eq:CSC}) and similar forms have been proposed in \cite{CMP4}, \cite{CMP0}, \cite{CMP2}, \cite{CMP}, \cite{CMP3}. They are based on MP or OMP, with the inner products between the residual and the columns of $\text{D}$ computed efficiently using convolutions with columns of $\text{D}_L$:
\[\textbf{b}_j=\text{D}_j^T\textbf{x}=\overleftrightarrow{\textbf{d}_j} \ast \textbf{x},\]
where $\overleftrightarrow{\textbf{d}_j}$ is the column-stacked horizontally and vertically flipped atom $\textbf{d}_j$. Thus, there is no need to store or use the explicit form of $\text{D}$ for matrix multiplication.

The computational complexity of most implementations of the convolutional matching pursuit is $O\left( N\text{log}\left( N\right)pk\right)$ (e.g. \cite{CMP4}, \cite{CMP3}), and the implementation in \cite{CMP} has a complexity of $O\left( Npk+np^2\right)$. Although $p$ is of the order of $n$ rather than $N$, as the atoms are local, convolutional sparse approximations typically have much larger values of the sparsity $k$ compared the patch sparsity $k_p$. This complexity is prohibitive, especially for large global images, which do not only increase the factor $N$ but also require a larger number of atoms for the representation (a larger $k$). Hence, later and more widely used methods surveyed in \cite{Survey2}, \cite{Survey3}, \cite{Survey1} use $\ell_1$ relaxation and minimize its unconstrained penalized Lagrangian form (similar to $P_1^\lambda$).

By imposing a constraint on the $\ell_0$ or $\ell_1$ norm of a global image, we can achieve a global sparse representation with a shift-invariant dictionary. However, in this global approach the selected atoms may be concentrated in some areas of the image, leaving other areas very sparse. In addition, previous works \cite{OptimallySparse}, \cite{GreedIsGood} have shown that pursuit algorithms are guaranteed to succeed as long as the $\ell_0$ ``norm'' is lower than a certain threshold. Therefore, a local pursuit method that is able to succeed in the sparser patches, which might cover most of the global image, could fail in its denser patches.

\subsection{The $P_{0,\infty}$ problem}
\begin{figure}[!t]
\centering
\includegraphics[width=5.0in]{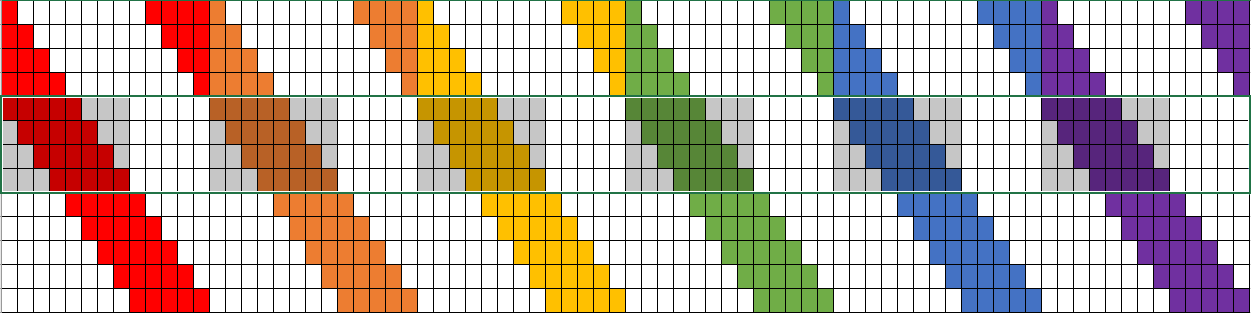}
\caption{A convolutional dictionary is the concatenation of matrices representing convolutions with different kernels (depicted in different colors). A stripe is a set of indices of columns whose contribution to a specific pixel is not trivially zero (marked in gray).}
\label{fig_stripe}
\end{figure}
A new prior for convolutional sparse coding was recently proposed in \cite{WorkingLocally}. Rather than minimizing the $\ell_0$ ``norm'' (the total number of nonzero coefficients in the convolutional representation), this model minimizes the $\ell_{0,\infty}$ group ``norm''. We will loosely follow the original definitions and notation of \cite{WorkingLocally}. The $P_{0,\infty}$ problem is defined as:
\begin{equation}
\left( P_{0,\infty} \right):~~~\underset{\boldsymbol{\alpha}}{\text{min}}  \left\Vert \boldsymbol{\alpha} \right\Vert_{0,\infty} ~s.t.~\textbf{x}=\text{D}\boldsymbol{\alpha}.
\label{eq:P0inf}
\end{equation}
Recall that the global dictionary $\text{D}$ is a concatenation of the convolution matrices of local atoms of size $n$, zero-padded to the size of the global signal $N$. A stripe $\Omega_i$ is defined as a set of $(2n-1)p$ indices of columns in $\text{D}$ whose values are not trivially zero at the row corresponding to the pixel $i$ ($1 \leq i \leq N$). Only nonzero coefficients of $\boldsymbol{\alpha}$ with these indices contribute to the value of the pixel $x_i$. Fig.~\ref{fig_stripe} demonstrates the concept of a convolutional dictionary and a stripe in it. The $\ell_{0,\infty}$ ``norm''\footnote{$\ell_{0,\infty}$ is also not actually a norm, but we shall nonetheless refer to it as a norm for the sake of brevity.} of the vector $\boldsymbol{\alpha}$, denoted $\left\Vert \boldsymbol{\alpha} \right\Vert_{0,\infty}$, is the number of nonzero coefficients in the densest stripe, or equivalently the maximum number of atoms contributing to any pixel. By limiting the sparsity of the densest stripe, we are effectively limiting the sparsity of all stripes, and therefore the number of overlaps of atoms in all pixels of the global representation. This new type of convolutional sparse coding allows representations with both a localized measure of sparsity and a shift invariant dictionary. This notion of sparsity is more intuitive and closer related to the original non-convolutional sparse coding problem in that for any given pixel in the image, there is a constraint on the number of atoms that may contribute to its value.

When modeling natural images, we allow some deviation from the model. As with $P_0^\epsilon$, instead of seeking a perfect reconstruction, we solve the $P_{0,\infty}^\epsilon$ problem:
\begin{equation}
\left( P_{0,\infty}^\epsilon \right):~\underset{\boldsymbol{\alpha}}{\text{min}}  \left\Vert \boldsymbol{\alpha} \right\Vert_{0,\infty} ~s.t.~\left\Vert \textbf{x} - \text{D}\boldsymbol{\alpha} \right\Vert_2^2 \leq \epsilon.
\label{eq:P0infeps}
\end{equation}

The work in \cite{WorkingLocally} provides guarantees for the success of the standard OMP and $\ell_1$ relaxation in the ideal and noisy regimes. It proposes optimization methods for solving an approximation of (\ref{eq:P0infeps}) using the Alternating Direction Method of Multipliers (ADMM) \cite{ADMM} to minimize the $\ell_1$ norm of the representation with additional penalties, which bias the solution towards a small $\ell_{0,\infty}$ ``norm''. In \cite{LocalProcessing}, a corresponding dictionary learning algorithm is proposed, which also minimizes an unconstrained Lagrangian with an $\ell_1$ norm penalty for the representation vector and an additional penalty for ensuring a small $\ell_{0,\infty}$ ``norm''. They refer to their method as \emph{slice-based convolutional sparse coding}. 

\subsection{Contribution}
Convolutional sparse coding with the $\ell_{0,\infty}$ prior has been recently proposed for natural images. However, the algorithms that have so far been introduced for solving it are relaxation based and only tackle the $\ell_{0,\infty}$ ``norm'' indirectly, relying on $\ell_1$ minimization and unconstrained optimization \cite{LocalProcessing}, \cite{WorkingLocally}.

In this work, we propose a novel greedy sparse coding scheme for the $\ell_{0,\infty}$ ``norm'' constrained minimization problem (\ref{eq:P0infeps}) and the corresponding dictionary learning problem\footnote{A code package which reproduces the experiments presented in this work is available at http://web.eng.tau.ac.il/$\sim$raja}. It allows direct control over the squared error or the $\ell_{0,\infty}$ sparsity, which enables incorporating prior knowledge of the sparsity when it is known. 
In the case of non-convolutional dictionary learning, greedy methods are often preferred for the sparse coding step because they are more computationally efficient \cite{GreedIsGood}. When solving (\ref{eq:DL1}), they allow designing a dictionary for a target sparsity.

Greedy strategies have so far been proposed for solving the standard convolutional sparse coding problem \cite{CMP4}, \cite{CMP0}, \cite{CMP2}, \cite{CMP}, \cite{CMP3}. Yet, they are not used often as they are computationally demanding. Our proposed greedy methods are able to efficiently solve the $\ell_{0,\infty}$ based convolutional sparse coding problem. They are efficient both in terms of computational complexity and memory requirement, and recover multiple sparse representation coefficients at the scale of the image after computing inner products with the columns of the local dictionary only once.

\section{A greedy approach to $P_{0,\infty}$}
\label{sectionPursuits}
\subsection{Group convolutional matching pursuit}
Recall the standard MP algorithm: (i) calculate inner products between the signal and all atoms; (ii) select the one with the largest absolute value and use the inner product as a coefficient; (iii) subtract the projection from the signal, creating a residual vector; and (iv) repeat steps (i)-(iii) with the residual as the new signal until a stopping condition is reached, such as a target sparsity.

In the case of the $P_{0,\infty}$ problem, we do not count the total number of selected atoms, but rather the number of their overlaps. After selecting the first atom as the one with the most significant inner product, we may add many more atoms without increasing the $\ell_{0,\infty}$ ``norm'', as long as they do not overlap. We create a representation that covers the entire image with non-overlapping atoms (Fig.~\ref{fig_hill_orig_cn}(a)) by sequentially selecting atoms with the most significant inner products excluding those which overlap atoms that have already been selected. Technically, we exclude atoms by zeroing their corresponding elements in the vector storing the inner products, thus, ensuring they will not be selected. We do so until it is not possible to add a new atom anywhere in the image without overlapping an existing atom. Until this point, the $\ell_{0,\infty}$ ``norm'' of the approximation equals one. Next, we subtract the current representation from the original signal, compute the inner products between the residual and the convolutional dictionary (using convolution operations), and create a second representation of non-overlapping atoms (Fig.~\ref{fig_hill_orig_cn}(b)). We continue stacking layers of non-overlapping atoms to the approximation until a sufficiently small error is attained if we target (\ref{eq:P0infeps}), or until a desired sparsity is reached if we solve another form of the problem:
\begin{equation}
\left( P_{0,\infty}^k \right):~~~\underset{\boldsymbol{\alpha}}{\text{min}} 
\left\Vert \textbf{x} - \text{D}\boldsymbol{\alpha} \right\Vert_2^2 ~s.t.~ \left\Vert \boldsymbol{\alpha} \right\Vert_{0,\infty} \leq k.
\label{eq:P0infk}
\end{equation}

The resulting algorithm is summarized in Algorithm \ref{algo:GCMP}, which we call Group Convolutional Matching Pursuit (GCMP). The inner loop adds the atoms with the most significant inner products to the representation, while excluding atoms that overlap those that have already been selected in the current layer. Thus, each iteration of the inner loop increments the $\ell_0$ ``norm'' while holding the $\ell_{0,\infty}$ ``norm'' fixed. Each iteration of the outer loop increments the $\ell_{0,\infty}$ norm. Notice that the number of times the inner products between the residual and the columns of the convolutional dictionary are calculated is equal to the $\ell_{0,\infty}$ norm, which is much smaller than the $\ell_0$ norm. This results from the fact that in each iteration of the inner loop, the added atoms do not overlap the ones selected so far, ensuring that the residual remains unchanged. Thus, there is no need to recompute the inner products with the updated residual until the next iteration of the outer loop.

\begin{algorithm}
\caption{Group Convolutional Matching Pursuit (GCMP)}
\begin{algorithmic}[0]
\State \textbf{Initialize:}~~$\boldsymbol{\alpha} \gets \textbf{0},~~~\textbf{r} \gets \textbf{x},~~~k \gets 0$
\While{$\left\Vert \textbf{r}  \right\Vert_2 > \epsilon$}
\State $\textbf{b} \gets \left[\begin{array}{cccc}
\left( \overleftrightarrow{\textbf{d}_1} \ast \textbf{r} \right)^T & \left( \overleftrightarrow{\textbf{d}_2} \ast \textbf{r} \right)^T & ... & \left( \overleftrightarrow{\textbf{d}_{p}} \ast \textbf{r} \right)^T
\end{array}\right]^T$
\While{$\underset{i}{\text{max}}~\left\{ \left\vert b_i \right\vert \right\} > 0$}
\State $i^* \gets \text{arg}~\underset{i}{\text{max}}~\left\{ \left\vert b_i \right\vert \right\}$
\State $\alpha_{i^*} \gets \alpha_{i^*} + b_{i^*}$
\For{$i \in \Omega_{i^*}$}{}
\State $b_i \gets 0$
\EndFor
\EndWhile
\State $\textbf{r} \gets \textbf{x} - \overset{p}{\underset{j=1}{\sum}} \textbf{d}_j \ast \boldsymbol{\alpha}_j$
\State $k \gets k+1$
\EndWhile
\end{algorithmic}
\label{algo:GCMP}
\end{algorithm}

Consequently, the order of complexity of such an algorithm is $O \left( N \text{log} \left( N \right) p \left\Vert \boldsymbol{\alpha} \right\Vert_{0,\infty} \right)$, which is much lower than the $O \left( N \text{log} \left( N \right) p \left\Vert \boldsymbol{\alpha} \right\Vert_{0} \right)$ required by the standard convolutional matching pursuit (e.g. \cite{CMP3}, \cite{CMP4}). If we use an implementation based on \cite{CMP}, the complexity becomes $O \left( N p \left\Vert \boldsymbol{\alpha} \right\Vert_{0,\infty} + np^2 \right)$, compared to $O \left( N p \left\Vert \boldsymbol{\alpha} \right\Vert_{0} + np^2 \right)$.

A useful inequality relates the $\ell_{0,\infty}$ sparsity to the overall $\ell_0$ sparsity. For a circular convolutional dictionary:
\begin{equation}
\left\Vert \boldsymbol{\alpha} \right\Vert_{0,\infty} \leq  \left\Vert \boldsymbol{\alpha} \right\Vert_0 \leq \frac{N}{n}\left\Vert \boldsymbol{\alpha} \right\Vert_{0,\infty}~,
\label{eq:ineqL0L0inf}
\end{equation}
and for a dictionary based on the linear (non-circular) convolution, $N$ is replaced by $N+n-1$. This inequality is easily understood by noticing that, per layer, the largest number of atoms that can cover the entire image without any overlaps occurs when they completely cover the image with no gaps between them.

\subsection{Group convolutional orthogonal matching pursuit}
A natural extension of our convolutional version of matching pursuit is an algorithm based on its orthogonal version, which we call Group Convolutional Orthogonal Matching Pursuit (GCOMP) and is summarized in Algorithm \ref{algo:GCOMP}. GCOMP differs from GCMP in that after every iteration of the outer loop, all the coefficients are updated by computing the orthogonal projection of the residual signal onto the set of atoms selected so far, denoted by $S$. Unfortunately, these cannot be computed using the convolution operation. Nevertheless, the full global dictionary does not need to be computed or stored, but only the columns corresponding to the indices in $S$. The matrix containing columns of $\text{D}$ with indices in $S$ is denoted by $\text{D}_S$. In each iteration of the outer loop, we update the representation coefficients by solving a least squares problem using the QR algorithm (orthogonal-triangular decomposition).

Notice that the matrix $D_s^T D_s$ is invertible. This can be seen by observing that as in regular OMP, in each iteration of GCOMP all the selected atoms are linearly independent of all the already selected ones. This happens because we constrain them not to overlap in the inner loop. In addition, after each iteration of the outer loop, we update the coefficients in the representation by least squares, which ensures that the residual in the next iteration of the outer loop is orthogonal to the current representation (similar to regular OMP). Notice also that the number of elements in the set $S$ is equal to the sparsity of a single representation of non-overlapping atoms, which according to inequality (\ref{eq:ineqL0L0inf}) is at most $\frac{N}{n}$. For high-fidelity representations, $\text{D}_S$ is computed more times, but the number of columns in it is bounded by inequality (\ref{eq:ineqL0L0inf}) every time.

\begin{algorithm}
\caption{Group Convolutional Orthogonal Matching Pursuit (GCOMP)}
\begin{algorithmic}[0]
\State \textbf{Initialize:}~~$\boldsymbol{\alpha} \gets \textbf{0},~~~\textbf{r} \gets \textbf{x},~~~k \gets 0,~~~S \gets \left\{ \right\}$
\While{$\left\Vert \textbf{r}  \right\Vert_2 > \epsilon$}
\State $\textbf{b} \gets \left[\begin{array}{cccc}
\left( \overleftrightarrow{\textbf{d}_1} \ast \textbf{r} \right)^T & \left( \overleftrightarrow{\textbf{d}_2} \ast \textbf{r} \right)^T & ... & \left( \overleftrightarrow{\textbf{d}_{p}} \ast \textbf{r} \right)^T
\end{array}\right]^T$
\While{$\underset{i}{\text{max}}~\left\{ \left\vert b_i \right\vert \right\} > 0$}
\State $i^* \gets \text{arg}~\underset{i}{\text{max}}~\left\{ \left\vert b_i \right\vert \right\}$
\State $S \gets S \cup i^*$
\For{$i \in \Omega_{i^*}$}{}
\State $b_i \gets 0$
\EndFor
\EndWhile
\State $\boldsymbol{\alpha}_S \gets \text{arg}~\underset{\boldsymbol{\alpha}_S}{\text{min}} \left\Vert \textbf{r} - \text{D}_S \boldsymbol{\alpha}_S \right\Vert_2^2$ \Comment{QR solver}
\State $\textbf{r} \gets \textbf{x} - \overset{p}{\underset{j=1}{\sum}} \textbf{d}_j \ast \boldsymbol{\alpha}_j$
\State $k \gets k+1$
\EndWhile
\end{algorithmic}
\label{algo:GCOMP}
\end{algorithm}

\subsection{Group convolutional thresholding}
A much simpler greedy algorithm based on the thresholding algorithm computes the inner products with the columns of the convolutional dictionary only once. Then, it selects those corresponding to the largest absolute values while excluding those that if added to the representation would violate the constraint on the $\ell_{0,\infty}$ norm. When there are no more atoms that can be selected without violating the constraint, all the coefficients are updated by computing the orthogonal projection of the signal onto the set of selected atoms, which we do by solving a least squares problem using the QR algorithm. The resulting strategy for approximating (\ref{eq:P0infk}) is summarized in Algorithm \ref{algo:GCT}. By modifying the stopping condition it may also approximate (\ref{eq:P0infeps}). We denote by $\left\Vert S \right\Vert_{0,\infty}$ the $\ell_{0,\infty}$ ``norm'' of any vector whose elements are nonzero at the indices in the set $S$.

\begin{algorithm}
\caption{Group Convolutional Thresholding (GCT)}
\begin{algorithmic}[0]
\State \textbf{Initialize:}~~$\boldsymbol{\alpha} \gets \textbf{0},~~~S \gets \left\{ \right\}$
\State $\textbf{b} \gets \left[\begin{array}{cccc}
\left( \overleftrightarrow{\textbf{d}_1} \ast \textbf{x} \right)^T & \left( \overleftrightarrow{\textbf{d}_2} \ast \textbf{x} \right)^T & ... & \left( \overleftrightarrow{\textbf{d}_{p}} \ast \textbf{x} \right)^T
\end{array}\right]^T$
\While{$\underset{i}{\text{max}}~\left\{ \left\vert b_i \right\vert \right\} > 0$}
\State $i^* \gets \text{arg}~\underset{i}{\text{max}}~\left\{ \left\vert b_i \right\vert \right\}$
\If{$\left\Vert S \cup i^* \right\Vert_{0,\infty} \leq k$}
\State $S \gets S \cup i^*$
\EndIf
\State $b_{i^*} \gets 0$
\EndWhile
\State $\boldsymbol{\alpha}_S \gets \text{arg}~\underset{\boldsymbol{\alpha}_S}{\text{min}} \left\Vert \textbf{x} - \text{D}_S \boldsymbol{\alpha}_S \right\Vert_2^2$ \Comment{QR solver}
\end{algorithmic}
\label{algo:GCT}
\end{algorithm}

In the case of exact recovery, $\textbf{x}=\text{D}\boldsymbol{\alpha}$, a performance guarantee for group convolutional thresholding (GCT) can be established using the mutual coherence of the dictionary and the $\ell_{0,\infty}$ norm.
If a solution $\boldsymbol{\alpha}$ exists obeying $\left\Vert \boldsymbol{\alpha} \right\Vert_{0,\infty} < \frac{1}{2} \left( \frac{\left\vert \alpha_{min} \right\vert}{\left\vert \alpha_{max} \right\vert} \frac{1}{\mu \left(\text{D}\right)} + 1 \right)$, then GCT is guaranteed to recover it exactly.
This guarantee is based on a result from \cite{CNNviaCSC}, where hard thresholding is used as a step in a forward pass through a multi-layer model. Note that the authors of \cite{CNNviaCSC} do not employ the least squares step and therefore their inequality only guarantees recovery of the support.

\subsection{Stagewise group convolutional orthogonal matching pursuit}
In GCOMP, we created representations with $\left\Vert \boldsymbol{\alpha} \right\Vert_{0,\infty} = k$ using $k$ iterations of the outer loop, each time computing the inner products and selecting the atoms in a way that increments $\left\Vert \boldsymbol{\alpha} \right\Vert_{0,\infty}$ by one. 
In GCT, we created the entire representation by computing the inner products between the signal and the atoms only once, and add atoms as long as $\left\Vert \boldsymbol{\alpha} \right\Vert_{0,\infty} \leq k$.

An alternative approach is to create the representation in stages, where at each stage we increment $\left\Vert \boldsymbol{\alpha} \right\Vert_{0,\infty}$ by some $1 \leq \Delta k \leq k$. When $\Delta k = 1$, this is equivalent to GCOMP, and when $\Delta k = k$, this is equivalent to GCT. When $1 < \Delta k < k$, at each outer iteration this method applies GCT with a sparsity of $\Delta k$, and repeats until reaching $\left\Vert \boldsymbol{\alpha} \right\Vert_{0,\infty} = k$, each time increasing $\left\Vert \boldsymbol{\alpha} \right\Vert_{0,\infty}$ by $\Delta k$.
This technique is summarized in Algorithm \ref{algo:StGCOMP}, which we call Stagewise GCOMP (after Stagewise OMP \cite{StagewiseOMP}). An alternative way to the usage of a constant increase in $\left\Vert \boldsymbol{\alpha} \right\Vert_{0,\infty}$ is to add all the atoms for which the amplitudes of their inner products with the current residual is larger than a certain threshold, as implemented in \cite{StagewiseOMP}.

\begin{algorithm}
\caption{Stagewise group convolutional OMP (StGCOMP)}
\begin{algorithmic}[0]
\State \textbf{Initialize:}~~$k \gets 0, ~~~\boldsymbol{\alpha} \gets \textbf{0},~~~S \gets \left\{ \right\}$
\Repeat
\State $\textbf{b} \gets \left[\begin{array}{cccc}
\left( \overleftrightarrow{\textbf{d}_1} \ast \left( \textbf{x} - \text{D}_S \boldsymbol{\alpha}_S \right) \right)^T & \left( \overleftrightarrow{\textbf{d}_2} \ast \left( \textbf{x} - \text{D}_S \boldsymbol{\alpha}_S \right) \right)^T & ... & \left( \overleftrightarrow{\textbf{d}_{p}} \ast \left( \textbf{x} - \text{D}_S \boldsymbol{\alpha}_S \right) \right)^T
\end{array}\right]^T$
\While{$\underset{i}{\text{max}}~\left\{ \left\vert b_i \right\vert \right\} > 0$}
\State $i^* \gets \text{arg}~\underset{i}{\text{max}}~\left\{ \left\vert b_i \right\vert \right\}$
\If{$\left\Vert S \cup i^* \right\Vert_{0,\infty} \leq \Delta k$}
\State $S \gets S \cup i^*$
\EndIf
\State $b_{i^*} \gets 0$
\EndWhile
\State $\boldsymbol{\alpha}_S \gets \boldsymbol{\alpha}_S +\text{arg}~\underset{\boldsymbol{\alpha}_S}{\text{min}} \left\Vert \textbf{x} - \text{D}_S \boldsymbol{\alpha}_S \right\Vert_2^2$ \Comment{QR solver}
\State $k \gets k + \Delta k$
\Until{reaching a desired value of $k$}
\end{algorithmic}
\label{algo:StGCOMP}
\end{algorithm}

\section{Convolutional Dictionary Learning}
Until now, we have assumed that the convolutional dictionary $\text{D}$ is given. Now, we turn to discuss some approaches for learning the set of convolutional kernels from examples. Let $\left\{ \textbf{x}^{(i)} \right\}_{i=1}^s$ be a training set of $s$ global signals and $\left\{ \textbf{d}_j \right\}_{j=1}^p$ be a set of $p$ local atoms. The dictionary is trained by optimizing both the sparse representations and the dictionary:
\begin{equation}
\underset{\left\{\boldsymbol{\alpha}^{(i)}\right\}_{i=1}^{s},\left\{\textbf{d}_j\right\}_{j=1}^{p}}{\text{min}} 
\overset{s}{\underset{i=1}{\sum}} \left\Vert \textbf{x}^{(i)} - \overset{p}{\underset{j=1}{\sum}}\textbf{d}_j \ast \boldsymbol{\alpha}_{j}^{(i)} \right\Vert_2^2 ~s.t.~ \left\Vert \boldsymbol{\alpha}^{(i)} \right\Vert_{0,\infty} \leq k,1 \leq i \leq s,
\label{eq:DL0inf}
\end{equation}
where $\boldsymbol{\alpha}^{(i)} \in \mathbb{R}^{Np \times 1}$ is a vector formed by vertically stacking $\boldsymbol{\alpha}_{j}^{(i)}$ for all $j$, and the constraint is on the $\ell_{0,\infty}$ ``norm'' of each $\boldsymbol{\alpha}^{(i)}$ separately.

We repeatedly alternate between a sparse coding step with a constraint on the $\ell_{0,\infty}$ norm, and a dictionary update step. The sparse coding problem is identical to (\ref{eq:P0infk}) and is solved by applying one of the pursuits introduced in Section \ref{sectionPursuits} (e.g. GCMP) to each of the $s$ global images separately (possibly in parallel). The dictionary update step minimizes the sum of total squared representation errors:
\begin{equation}
\underset{\left\{\textbf{d}_j\right\}}{\text{min}}~\overset{s}{\underset{i=1}{\sum}}
\left\Vert \textbf{x}^{(i)} - \overset{p}{\underset{j=1}{\sum}}\textbf{d}_j \ast \boldsymbol{\alpha}_{j}^{(i)} \right\Vert_2^2.
\label{eq:Dupdate}
\end{equation}

We experimented with two different optimization methods for the dictionary update step, which we describe next. Additional dictionary learning methods have recently been surveyed in \cite{Survey3}.

\subsection{Convolutional method of optimal directions}
First, we use the fact that the convolution operation can be constructed as a matrix multiplication to rewrite the convolution in (\ref{eq:DL0inf}) as a matrix multiplication:
\[
\underset{\left\{\textbf{d}_j\right\}_{j=1}^{p}}{\text{min}}~\overset{s}{\underset{i=1}{\sum}}
\left\Vert \textbf{x}^{(i)} - \overset{p}{\underset{j=1}{\sum}}\text{A}_{j}^{(i)}\textbf{d}_j  \right\Vert_2^2 ,\]
where $\text{A}_{j}^{(i)}$ is the convolution matrix of $\boldsymbol{\alpha}_{j}^{(i)}$. We define the matrix $\text{A}^{(i)}$ as the horizontal concatenation of the convolution matrices of the sparse representation of signal $i$ for all $j$ 
$\text{A}^{(i)}=\left[\begin{array}{cccc}\text{A}_{1}^{(i)} & \text{A}_{2}^{(i)} & ... & \text{A}_{p}^{(i)}\end{array}\right]$, and the matrix $\text{A} \in \mathbb{R}^{Ns \times p}$ by stacking the matrices $\text{A}^{(i)}$ vertically. We define the vector $\textbf{d} \in \mathbb{R}^{np \times 1}$ as a vertical concatenation of the local dictionary $\textbf{d}=\left[\begin{array}{cccc}\textbf{d}_{1}^T & \textbf{d}_{2}^T & ... & \textbf{d}_{p}^T\end{array}\right]^T$, and the matrix $\textbf{x} \in \mathbb{R}^{Ns \times 1}$ as the vertical concatenation of the training signals. Thus, (\ref{eq:Dupdate}) can be rewritten as minimization of a squared Frobenius norm with respect to a single vector:
\begin{equation} \underset{\textbf{d}}{\text{min}}~
\left\Vert \textbf{x} - \text{A}\textbf{d}  \right\Vert_F^2.
\end{equation}
An analytical solution can be found by taking the gradient and setting it to zero, yet its computational complexity is prohibitive. Nevertheless, this is an unconstrained convex minimization, which can be solved by a variety of numerical methods. The gradients only require the computation of $\text{A}^T \textbf{x}$ and $\text{A}^T \text{A} \textbf{d}$, which can both be computed efficiently using convolution operations: the cost of computing $\text{A}^T \textbf{x}$ is $O \left( spN\text{log}\left( N\right)\right)$, and the cost of computing $\text{A}^T \text{A} \textbf{d}$ is $O \left( sp^2N\text{log}\left( N\right)\right)$.

We choose to use Conjugate Gradient Least Squares (CGLS) with convolution operations. If we loop through all $mnp$ conjugate directions, the total cost of each dictionary update step per image is $O \left( p^2nN\text{log}\left( N\right) + p^3nN\text{log}\left( N\right)\right) \approx O \left( p^3nN\text{log}\left( N\right)\right)$, as typically $k \ll p$. However, it has been shown \cite{Survey1} that early stopping of the CGLS step (e.g., with an error of $10^{-3}$) is sufficient for reliable convergence. This greatly reduces the complexity to $O \left( p^2qN\text{log}\left( N\right)\right)$, where $q$ is the number of inner iterations in the CGLS step ($q \ll np$).

After updating the dictionary, we go back to sparse coding with the updated dictionary (e.g., using GCMP). Then, the next dictionary update occurs using the new values of $\text{A}$, and we repeat until convergence in the total representation error. The resulting dictionary learning algorithm, which is a convolutional version of MOD \cite{MOD}, is summarized in Algorithm \ref{algo:CMOD}. Notice that the constraint on the $\ell_{0,\infty}$ ``norm'' is enforced in the sparse coding step, and each dictionary update step optimizes the dictionary given the current representations $\text{A}$. 

\begin{algorithm}
\caption{Convolutional MOD}
\begin{algorithmic}[0]
\State Initialize \textbf{d} to some initial dictionary, column-stacked.
\Repeat
\For{$1 \leq i \leq s$}
\State $\boldsymbol{\alpha}^{(i)}\gets\text{arg}~\underset{\left\{\boldsymbol{\alpha}_i\right\}}{\text{min}}~
\left\Vert \textbf{x}^{(i)} - \overset{p}{\underset{j=1}{\sum}}\textbf{d}_j \ast \boldsymbol{\alpha}_{j}^{(i)} \right\Vert_2^2~~s.t.~~\left\Vert \boldsymbol{\alpha}^{(i)} \right\Vert_{0,\infty} \leq k$  \Comment{using GCMP}
\EndFor
\State $\textbf{d} \gets \text{arg}~\underset{\textbf{d}}{\text{min}}~\left\Vert \textbf{x} - \text{A}\textbf{d}  \right\Vert_F^2$ \Comment{using CGLS}
\For{$1 \leq j \leq p$
\State $\textbf{d}_j \gets \frac{\textbf{d}_j}{\left\Vert \textbf{d}_j \right\Vert_2}$
\EndFor}
\Until{convergence of $\left\Vert \textbf{x} - \text{A}\textbf{d}  \right\Vert_F^2$}
\end{algorithmic}
\label{algo:CMOD}
\end{algorithm}

\subsection{Convolutional block-coordinate descent}
Updating the whole dictionary at each iteration may become computationally challenging for large images or large dictionaries, even when using a gradient based optimizer with convolutional operations. Alternatively, we can update the atoms one at a time (as in K-SVD), which is equivalent to block-coordinate descent optimization. In this case, the dictionary update step loops through all atoms one at a time (possibly in parallel), each time minimizing the objective by a single atom, $\textbf{d}_{j_0}$:
\begin{equation}
\underset{\textbf{d}_{j_0}}{\text{min}} 
\overset{s}{\underset{i=1}{\sum}} \left\Vert \textbf{x}^{(i)} - \overset{p}{\underset{j=1}{\sum}}\textbf{d}_j \ast \boldsymbol{\alpha}_{j}^{(i)} \right\Vert_2^2 = \underset{\textbf{d}_{j_0}}{\text{min}} 
\overset{s}{\underset{i=1}{\sum}} \left\Vert \textbf{e}_{j_0}^{(i)} - \textbf{d}_{j_0} \ast \boldsymbol{\alpha}_{j_0}^{(i)} \right\Vert_2^2~,
\end{equation}
where $\textbf{e}_{j_0}^{(i)}=\textbf{x}^{(i)}-\underset{j \neq j_0}{\sum}\textbf{d}_j \ast \boldsymbol{\alpha}_{j}^{(i)}$. The same solver from Algorithm \ref{algo:CMOD}, convolutional CGLS, can be used here with $\textbf{d}$ replaced by $\textbf{d}_{j_0}$, $\textbf{x}^{(i)}$ replaced by $\textbf{e}_{j_0}^{(i)}$ and $\text{A}^{(i)}$ replaced by $\text{A}_{j_0}^{(i)}$. The computation cost becomes $O \left( pnN\text{log}\left( N\right) + p^2nN\text{log}\left( N\right)\right) \approx O \left( p^2nN\text{log}\left( N\right)\right)$ per iteration or $O \left( pqN\text{log}\left( N\right)\right)$ with early stopping, which is significantly lower than the computational cost of Algorithm \ref{algo:CMOD}.
The resulting algorithm is summarized in Algorithm \ref{algo:CBCD}.

\begin{algorithm}
\caption{Convolutional BCD}
\begin{algorithmic}[0]
\State Initialize $\left\{ \textbf{d} \right\}_{j=1}^p$ to some initial dictionary.
\Repeat
\For{$1 \leq i \leq s$}
\State $\boldsymbol{\alpha}^{(i)}\gets\text{arg}~\underset{\boldsymbol{\alpha}^{(i)}}{\text{min}}~\left\Vert \textbf{x}^{(i)} - \overset{p}{\underset{j=1}{\sum}}\textbf{d}_j \ast \boldsymbol{\alpha}_{j}^{(i)} \right\Vert_2^2~~s.t.~~\left\Vert \boldsymbol{\alpha}^{(i)} \right\Vert_{0,\infty} \leq k$  \Comment{using GCMP}
\EndFor
\For{$1 \leq j \leq p$}
\For{$1 \leq i \leq s$}
\State $\textbf{e}_{j}^{(i)}=\textbf{x}^{(i)}-\underset{j' \neq j}{\sum}\textbf{d}_{j'} \ast \boldsymbol{\alpha}_{j'}^{(i)}$
\EndFor
\State $\textbf{d}_j \gets \text{arg}~\underset{\textbf{d}_{j}}{\text{min}}
\overset{s}{\underset{i=1}{\sum}} \left\Vert \textbf{e}_{j}^{(i)} - \textbf{d}_{j} \ast \boldsymbol{\alpha}_{j}^{(i)} \right\Vert_2^2$ \Comment{using CGLS}
\State $\textbf{d}_j \gets \frac{\textbf{d}_j}{\left\Vert \textbf{d}_j \right\Vert_2}$
\EndFor
\Until{convergence of $\left\Vert \textbf{x} - \text{A}\textbf{d}  \right\Vert_F^2$}
\end{algorithmic}
\label{algo:CBCD}
\end{algorithm}

\section{Computational complexity}
Table \ref{tableComplexity} summarizes the computational complexity of our sparse coding and dictionary learning methods compared to the methods referenced in previous sections. The complexities are given for a single signal, and the dictionary learning complexity is for a single dictionary update iteration. For convenience, we summarize here the notations used in the table. $N$ is the dimension of the global signal and $n$ is the dimensionality of the local dictionary (in patch-based methods this is the number of pixels in each patch, and in convolutional sparse coding this is the size of each filter). $p$ is the number of atoms in the dictionary (in convolutional sparse coding, this is the number of filters). In global sparse coding, $p>N$ for an overcomplete dictionary, and in patch based methods $p>n$ for an overcomplete dictionary. In convolutional sparse coding, often $p>n$. $k_p$ is the sparsity in patch based methods. In slice-based convolutional sparse coding and in our method, $k$ is the $\ell_{0,\infty}$ sparsity, and in $\ell_0$ convolutional matching pursuit it is the sparsity of the global signal (and is therefore larger). $q$ is the number of inner iterations required by some solvers. The computational complexity of sparse coding of an image using the method in \cite{Heide}, which is the same as its dictionary update step, is $O\left(qpN+qpN\text{log}\left(N\right)\right) \approx O\left(qpN\text{log}\left(N\right)\right)$, where $q$ is the number of inner iterations in their ADMM \cite{ADMM} solver. The work in \cite{Heide} used $q=10$ and their dictionary learning converged after about 13 outer iterations. Our method has inner iterations in the conjugate gradient dictionary update step. However, we usually do not loop through all conjugate directions and instead stop at some error tolerance.

\renewcommand{\arraystretch}{1.25}
\begin{table}
\centering
\small
\caption{Computational complexity}
$
\begin{array}{|c|c|c|} \hline
\text{Method} & ~\text{Sparse coding}~ & ~\text{Dictionary update}~  \\ \hline
\text{Global OMP + MOD} & O\left( Npk\right) & O\left( p^2+p^3\right) \\
\text{Global OMP + K-SVD} & O\left( Npk\right) & O\left( Np+Nk+pk\right) \\
\text{Patch-based OMP + MOD} & O\left( Nnpk\right) & O\left( Np^2+p^3\right) \\
\text{Patch-based OMP + K-SVD} & O\left( Nnpk\right) & O\left( np^2+Nnk+Npk\right) \\
\ell_0 \text{ based CSC \cite{CMP4}, \cite{CMP2}} &  O\left( N\text{log}\left( N\right)pk\right) & - \\
\ell_0 \text{ based CSC \cite{CMP}} &  O\left( Npk+np^2\right) & - \\
\ell_1 \text{ based CSC \cite{Heide}} &  O\left( N\text{log}\left( N\right)pq + Npq\right) & O\left( N\text{log}\left( N\right)pq + Npq \right) \\
\text{Slice based CSC \cite{LocalProcessing}} & O\left( Nnp + N\left( k^3+pk^2\right) \right) & O\left( np^2 + Nk \left( n+p\right) \right) \\
\text{Our method: GCMP + CBCD} & O\left( N\text{log}\left( N\right)pk \right) & O\left( N\text{log}\left( N\right)pq \right) \\ \hline
\end{array}
$
\label{tableComplexity}
\end{table}

\section{Methods}
Previous works on $\ell_{0,\infty}$ based convolutional sparse coding \cite{LocalProcessing} have demonstrated its applicability in image processing applications. In a similar way, we show next how we may adapt our greedy algorithms to such tasks, namely image inpainting and salt-and-pepper noise removal. In the supplementary material, we present an adaptation to another application: texture and cartoon separation.

\subsection{Inpainting}
\label{sectionInpainting}
Inpainting is the task of recovering an image from a corrupted version of it, which has missing pixels. We shall assume that the corrupted image, denoted $\textbf{y}$, holds the value zero in the corrupted pixels, whose locations are known, and is identical to the original image in all other pixels. We assume a convolutional sparse representation of the original image:
\[ \textbf{y}=\text{C}\textbf{x}=\text{CD}\boldsymbol{\alpha}=\text{C}\overset{p}{\underset{j=1}{\sum}} \textbf{d}_j \ast \boldsymbol{\alpha}_j ,\]
where $\text{C}$, referred to as the subsampling matrix, is a diagonal binary matrix with elements $c_{ii}=0$ for a corrupted pixel $i$, $c_{ii}=1$ for an uncorrupted pixel $i$, and $c_{ij}=0$ for $i \neq j$. We find the sparse representation of the original image by solving the $P_{0,\infty}^k$ problem with the global dictionary $\text{CD}$:
\begin{equation}
\underset{\boldsymbol{\alpha}}{\text{min}} 
\left\Vert \textbf{y} - \text{CD} \boldsymbol{\alpha} \right\Vert_2^2 ~s.t.~ \left\Vert \boldsymbol{\alpha} \right\Vert_{0,\infty} \leq k.
\end{equation}
The inner products between the corrupted image and the columns of \text{CD} are the same as with the columns of \text{D} due to \text{C} being symmetric and idempotent:
$ \textbf{b}=\left(\text{CD}\right)^T\textbf{y} = \text{D}^T\text{C}^T\text{C}\textbf{x}=\text{D}^T\text{C}\textbf{x} = \text{D}^T \textbf{y}$. Thus, these inner products can still be computed efficiently using convolution operations. The only effect \text{C} has on the pursuit algorithms is the computation of the residual, which becomes:
$\textbf{r} \gets \textbf{y} - \text{C}\overset{p}{\underset{j=1}{\sum}} \textbf{d}_j \ast \boldsymbol{\alpha}_j$. After computing the sparse representation, we estimate the original image as:
\[ \hat{\textbf{x}} = \text{D}\boldsymbol{\alpha} = \overset{p}{\underset{j=1}{\sum}} \textbf{d}_j \ast \boldsymbol{\alpha}_j. \]

Training of the dictionary on the corrupted image \cite{Heide}, \cite{ImageSpecific} can be done by solving the following minimization problem:
\begin{equation}
\underset{\boldsymbol{\alpha},\left\{\textbf{d}_j\right\}_{j=1}^p}{\text{min}} 
\left\Vert \textbf{y} - \text{C}\overset{p}{\underset{j=1}{\sum}}\textbf{d}_j \ast \boldsymbol{\alpha}_j \right\Vert_2^2 ~s.t.~ \left\Vert \boldsymbol{\alpha} \right\Vert_{0,\infty} \leq k.
\end{equation}
Following the steps of the dictionary learning algorithms, the dictionary update step can be computed by using CGLS to solve
\[ \textbf{d} \gets \text{arg}~\underset{\textbf{d}}{\text{min}}~\left\Vert \textbf{x} - \text{CA}\textbf{d}  \right\Vert_F^2, \]
when updating the whole dictionary at once (Convolutional MOD), or
\[\textbf{d}_j \gets \text{arg}~\underset{\textbf{d}_{j}}{\text{min}} 
\overset{s}{\underset{i=1}{\sum}} \left\Vert \text{C}\textbf{e}_{j}^{(i)} - \text{C} \textbf{d}_{j} \ast \boldsymbol{\alpha}_{j}^{(i)} \right\Vert_2^2, \]
when looping through all indices $j$ (Convolutional BCD). Because $\text{C}\text{A}$ is rank-deficient and the Hessian $\text{A}^T\text{C}\text{A}$ is ill-conditioned, we do not seek a full minimization at each iteration, but rather take a single step of gradient descent as the dictionary update. In order to ensure that all atoms are updated, we use convolutional BCD with a single step of gradient descent for each atom before recomputing the sparse representation. The same solver from Algorithm \ref{algo:CBCD} may be used with these adjustments and a step size $\gamma$. Looping through $1 \leq j \leq p$, the update of each atom reads as:
\[
\textbf{d}_j \gets \textbf{d}_j - \gamma \frac{\partial}{\partial \textbf{d}_j} \left\Vert \text{C} \left( \textbf{e}_{j} - \textbf{d}_{j} \ast \boldsymbol{\alpha}_{j} \right) \right\Vert_2^2 .
\]

\subsection{Salt-and-pepper noise removal}
Salt and pepper refers to noise that affects an image by turning some of the pixels into black or white (minimum and maximum gray levels). We do not assume knowledge of the noise mask, i.e. whether each pixel is corrupted or not. Rather, we rely on the prior knowledge that the image is sparse (in the $\ell_{0,\infty}$ sense) in a dictionary $\left\{\textbf{d}_j\right\}_{j=1}^{p}$, and that the noise is sparse in a noise dictionary which contains a single atom, the unit impulse:
$
\textbf{d}_0 = \left[\begin{array}{ccccc} 1 & 0 & 0 & ... & 0 \end{array} \right]^T.
$
Thus, for each noisy image, we solve:
\begin{equation}
\underset{\boldsymbol{\alpha}, \boldsymbol{\alpha}_0}{\text{min}} ~ \left\Vert \textbf{x} - \overset{p}{\underset{j=1}{\sum}} \textbf{d}_j \ast \boldsymbol{\alpha}_j - \textbf{d}_0 \ast \boldsymbol{\alpha}_0 \right\Vert_2^2 ~~s.t.~~ \left\Vert \boldsymbol{\alpha} \right\Vert_{0,\infty} \leq k,~ \left\Vert \boldsymbol{\alpha}_0 \right\Vert_{0,\infty} \leq k_{\text{noise}} ,
\end{equation}
by alternating between two steps: (i) convolutional sparse coding using the dictionary $\left\{\textbf{d}_j\right\}_{j=0}^{p}$, which includes the additional noise atom, and zeroing the coefficients of the noise atom in the reconstruction; (ii) convolutional sparse coding of the residual using only the noise atom, for an estimate of the noise. Then, we repeat step (i) for the image with the noise estimate subtracted from it. We repeatedly alternate between the two steps, each time increasing the $\ell_{0,\infty}$ ``norm'' until reaching the target sparsities for the image and the noise. Thus, the noisy image is separated into a text component and an impulse noise component. An exact description of the method appears in Algorithm \ref{algo:SNP}.

\begin{algorithm}
\caption{Salt-and-pepper noise removal}
\begin{algorithmic}[0]
\State \textbf{Initialize:}~~$\left\{\boldsymbol{\alpha}_j\right\}_{j=0}^p \gets \textbf{0},~~~~k_1 \gets 1, k_2 \gets 1$
\While{$k_1 \leq k$ and $k_2 \leq k_{\text{noise}}$}
\State $ \left\{\boldsymbol{\alpha}_j\right\}_{j=1}^p \gets \text{arg}~\underset{\left\{\boldsymbol{\alpha}_j\right\}_{j=1}^p}{\text{min}} 
\left\Vert \left( \textbf{x} - \textbf{d}_0 \ast \boldsymbol{\alpha}_0 \right) - \overset{p}{\underset{j=1}{\sum}}\textbf{d}_j \ast \boldsymbol{\alpha}_j \right\Vert_2^2~~s.t.~~\left\Vert 
\left[\begin{array}{c} \boldsymbol{\alpha}_1 \\ \vdots \\ \boldsymbol{\alpha}_p \end{array} \right] \right\Vert_{0,\infty} \leq k~$
\State $ \boldsymbol{\alpha}_0 \gets \text{arg}~\underset{\boldsymbol{\alpha}_0}{\text{min}} 
\left\Vert \left( \textbf{x} - \overset{p}{\underset{j=1}{\sum}}\textbf{d}_j \ast \boldsymbol{\alpha}_j \right) - \textbf{d}_0 \ast \boldsymbol{\alpha}_0 \right\Vert_2^2~~s.t.~~\left\Vert \boldsymbol{\alpha}_0 \right\Vert_{0,\infty} \leq k_{\text{noise}}~$
\State $k_1 \gets k_1+1$
\State $k_2 \gets k_2+1$
\EndWhile
\end{algorithmic}
\label{algo:SNP}
\end{algorithm}

\section{Experiments}
We turn now to evaluate our proposed strategy. We first perform an evaluation of the different components in the proposed approach and then compare it to other methods for inpainting and impulse noise removal. We use two types of data in our experiments: natural images with local contrast normalization, which is a very common type of data used for testing convolutional sparse coding \cite{Heide}, and text images.

For the latter, we use scanned pages from a book, Aristotle's \emph{Nicomachean Ethics}, taken from The Internet Archive (www.archive.org). We use the first 16 pages as the training set and the next 16 pages (pages 17-32) as the test set. The size of all images is $497 \times 383$ and gray levels are normalized to the range $\left[0,1\right]$, where zero is black and one is white. 
In the case of black-on-white text images, the background can be assumed to be white. As typically zero represents the black gray-level, we invert the gray-levels of the dataset so that the text is white-on-black. This way, the background is black (contains zeros) and not white, which is consistent with the fact that the gaps between atoms are zeros in GCMP, and no atoms are required to approximate them. At the end, we invert the result to a black-on-white image.

\subsection{Accuracy vs. sparsity}
\label{sec:acc_vs_sparsity}

\begin{figure}[!b]
\captionsetup[subfigure]{justification=centering,font=scriptsize}
    \centering
  \subfloat[$\left\Vert \boldsymbol{\alpha} \right\Vert_{0,\infty}=1$]{
       \begin{tikzpicture}[spy using outlines={circle,yellow,magnification=5,size=1.5cm, connect spies}]
\node {\pgfimage[interpolate=true,width=0.21\linewidth]{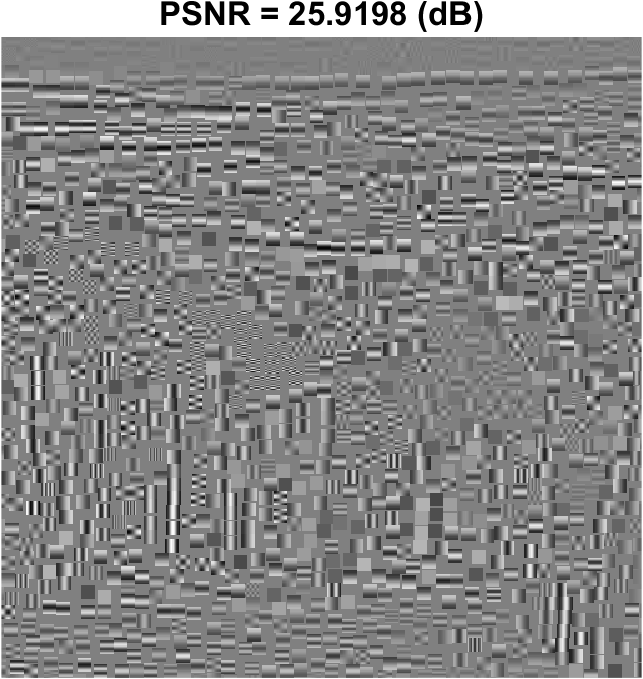}};
\spy on (0,0.4) in node [left] at (1.75,0.75);
\end{tikzpicture}
       }
    \label{hill_orig_cn_a} 
  \subfloat[$\left\Vert \boldsymbol{\alpha} \right\Vert_{0,\infty} = 2$]{
               \begin{tikzpicture}[spy using outlines={circle,yellow,magnification=5,size=1.5cm, connect spies}]
\node {\pgfimage[interpolate=true,width=0.21\linewidth]{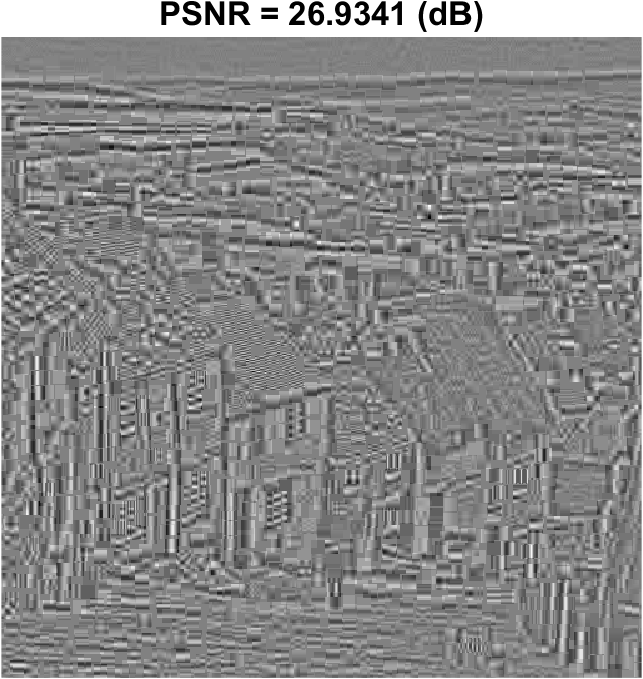}};
\spy on (0,0.4) in node [left] at (1.75,0.75);
\end{tikzpicture}
        }
    \label{hill_orig_cn_b}
  \subfloat[$\left\Vert \boldsymbol{\alpha} \right\Vert_{0,\infty} = 8$]{
               \begin{tikzpicture}[spy using outlines={circle,yellow,magnification=5,size=1.5cm, connect spies}]
\node {\pgfimage[interpolate=true,width=0.21\linewidth]{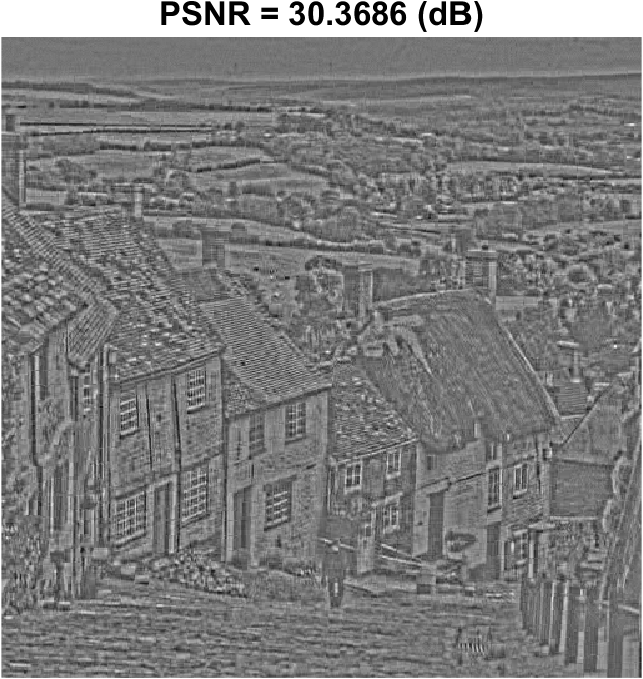}};
\spy on (0,0.4) in node [left] at (1.75,0.75);
\end{tikzpicture}
        }
    \label{hill_orig_cn_c} 
  \subfloat[$\left\Vert \boldsymbol{\alpha} \right\Vert_{0,\infty} = 32$]{
               \begin{tikzpicture}[spy using outlines={circle,yellow,magnification=5,size=1.5cm, connect spies}]
\node {\pgfimage[interpolate=true,width=0.21\linewidth]{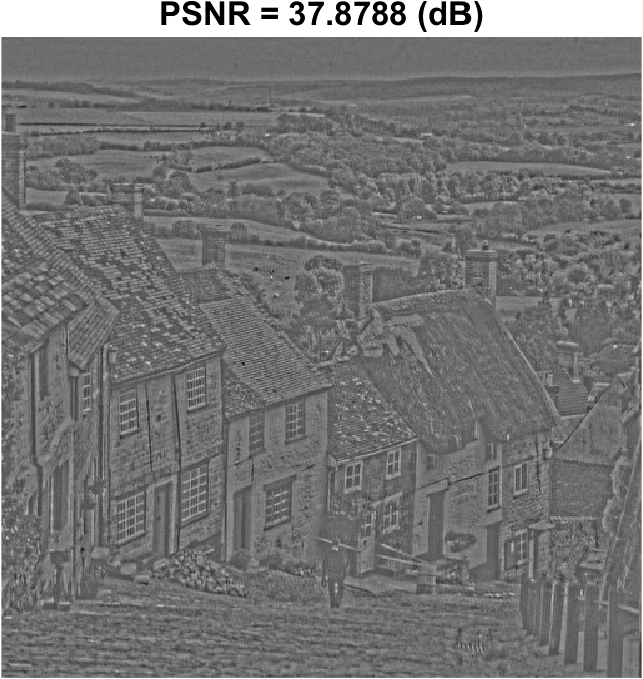}};
\spy on (0,0.4) in node [left] at (1.75,0.75);
\end{tikzpicture}
        }
    \label{hill_orig_cn_d}
  \subfloat[$\left\Vert \boldsymbol{\alpha} \right\Vert_{0,\infty}=1$]{
       \includegraphics[width=0.235\linewidth]{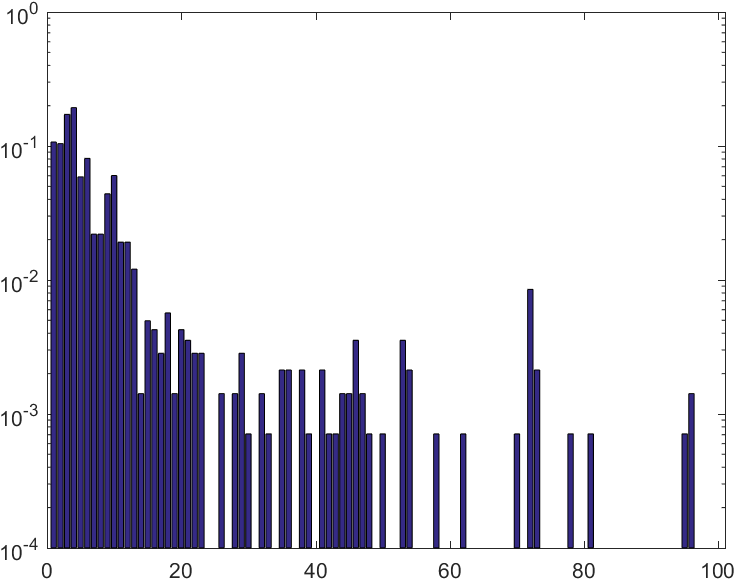}}
    \label{hill_orig_cn_e} 
  \subfloat[$\left\Vert \boldsymbol{\alpha} \right\Vert_{0,\infty}=2$]{
        \includegraphics[width=0.235\linewidth]{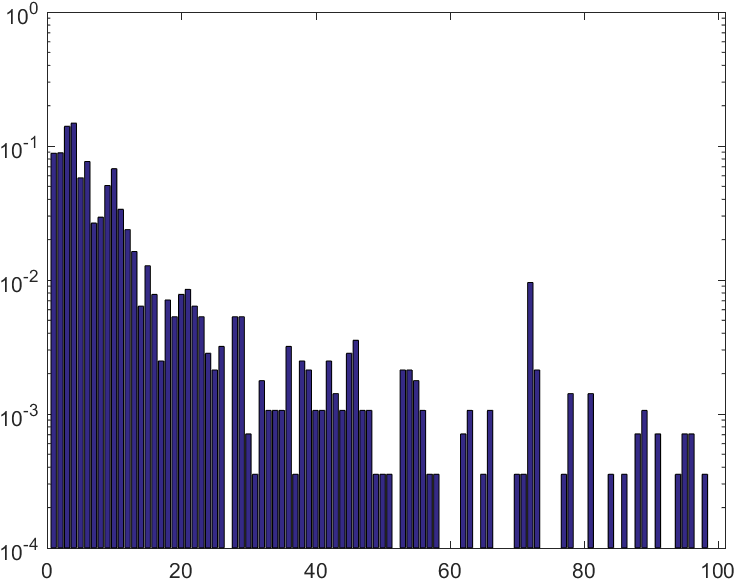}}
    \label{hill_orig_cn_f} 
  \subfloat[$\left\Vert \boldsymbol{\alpha} \right\Vert_{0,\infty}=8$]{
        \includegraphics[width=0.235\linewidth]{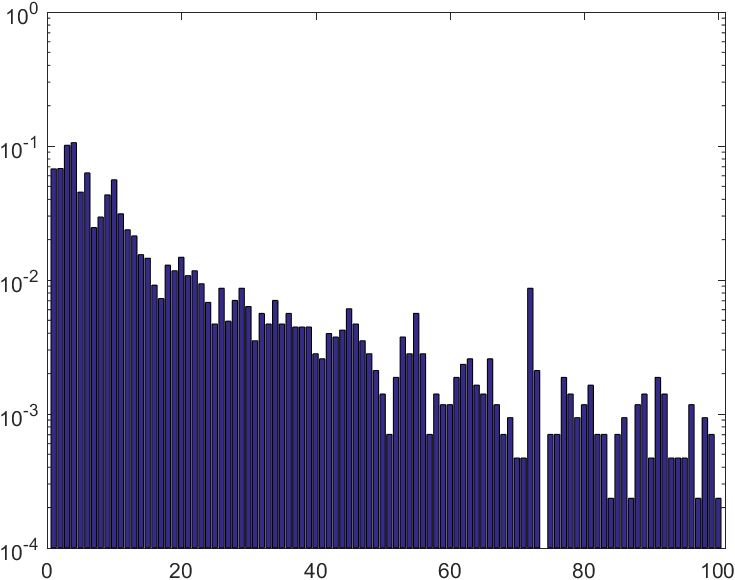}}
    \label{hill_orig_cn_g} 
  \subfloat[$\left\Vert \boldsymbol{\alpha} \right\Vert_{0,\infty}=32$]{
        \includegraphics[width=0.235\linewidth]{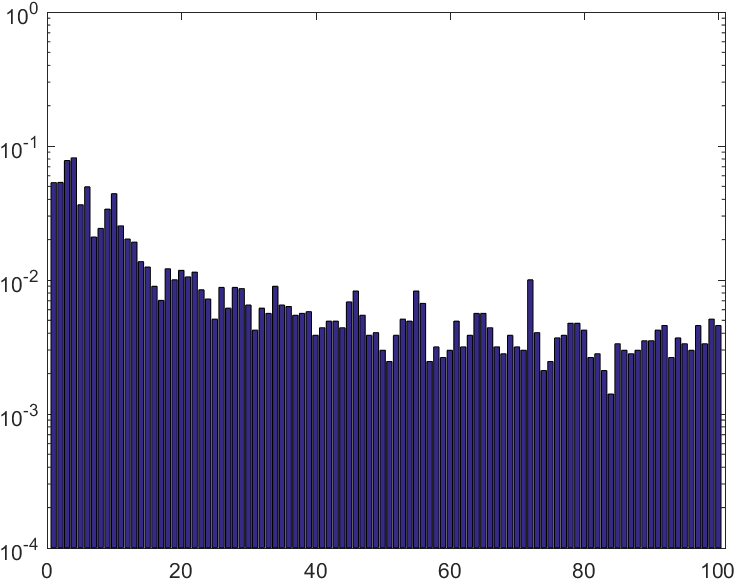}}
    \label{hill_orig_cn_h}
  \caption{(a)-(d) Approximations of the \emph{hill} image with local contrast normalization using GCMP for several $\ell_{0,\infty}$ sparsities and their PSNR values; (e)-(h) histograms of atoms selected for each sparsity (sorted in increasing spatial frequency).}
  \label{fig_hill_orig_cn}
\end{figure}

To illustrate how our proposed algorithm reconstructs the representation of an image, we apply GCMP (Algorithm \ref{algo:GCMP}) to 11 standard test images (taken from \cite{LocalProcessing}) after contrast normalization using an undercomplete Discrete Cosine Transform (DCT) dictionary with 100 atoms of size $11\times11$. Fig.~\ref{fig_hill_orig_cn}(a)-(d) show the reconstructed \emph{hill} image from representations with several values of $\left\Vert \boldsymbol{\alpha} \right\Vert_{0,\infty}$ (namely, 1, 2, 8 and 32) and their PSNR values. In addition, we plotted the histograms of the number of times each atom was selected, normalized by the $\left\Vert \boldsymbol{\alpha} \right\Vert_0$ sparsity (Fig.~\ref{fig_hill_orig_cn}(e)-(h)). The dictionary indices in the horizontal axis are sorted in an increasing spatial frequency. Notice how low frequency atoms (representing cartoon) are preferred in the first iterations, while the ones with high frequencies (containing texture details) are selected for larger values of the $\ell_{0,\infty}$ ``norm''. When we have performed the same visualization for images without local contrast normalization, we have found that a larger portion of atoms are dedicated to the DC component (the first bin in the histogram is much higher), which may explain the tendency of using local contrast normalized images in convolutional sparse coding experiments.

\begin{figure}[!b]
\captionsetup[subfigure]{justification=centering,font=scriptsize}
    \centering
  \subfloat[$\left\Vert \boldsymbol{\alpha} \right\Vert_{0,\infty}=1$ $\text{PSNR=17.13(dB)}$]{
                      \begin{tikzpicture}[spy using outlines={circle,yellow,magnification=5,size=1.5cm, connect spies}]
\node {\pgfimage[interpolate=true,width=0.21\linewidth]{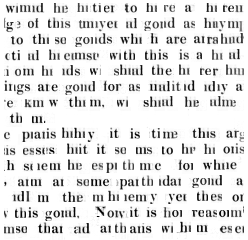}};
\spy on (0.165,0.655) in node [left] at (1.75,0.75);
\end{tikzpicture}
       }
    \label{textdict1a} 
  \subfloat[$\left\Vert \boldsymbol{\alpha} \right\Vert_{0,\infty} = 2$ $\text{PSNR=20.40	(dB)}$]{
                              \begin{tikzpicture}[spy using outlines={circle,yellow,magnification=5,size=1.5cm, connect spies}]
\node {\pgfimage[interpolate=true,width=0.21\linewidth]{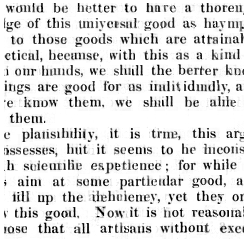}};
\spy on (0.165,0.655) in node [left] at (1.75,0.75);
\end{tikzpicture}
        }
    \label{textdict1b}
  \subfloat[$\left\Vert \boldsymbol{\alpha} \right\Vert_{0,\infty} = 10$ $\text{PSNR=26.01(dB)}$]{
        \begin{tikzpicture}[spy using outlines={circle,yellow,magnification=5,size=1.5cm, connect spies}]
\node {\pgfimage[interpolate=true,width=0.21\linewidth]{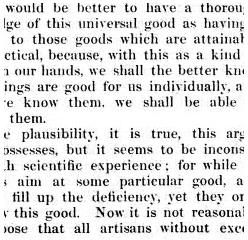}};
\spy on (0.165,0.655) in node [left] at (1.75,0.75);
\end{tikzpicture}
        }
    \label{textdict1c} 
  \subfloat[$\left\Vert \boldsymbol{\alpha} \right\Vert_{0,\infty} = 20$ $\text{PSNR=29.29(dB)}$]{
        \begin{tikzpicture}[spy using outlines={circle,yellow,magnification=5,size=1.5cm, connect spies}]
\node {\pgfimage[interpolate=true,width=0.21\linewidth]{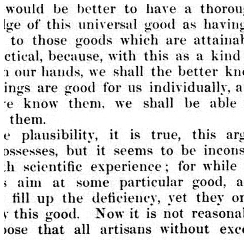}};
\spy on (0.165,0.655) in node [left] at (1.75,0.75);
\end{tikzpicture}
        }
    \label{textdict1d}
  \caption{Approximations of a text image using GCMP for several $\ell_{0,\infty}$ sparsities and their PSNR values. The dictionary was trained on text images.}
  \label{fig_textdict}
\end{figure}

Next, we train a dictionary of 100 atoms of size $11 \times 11$ on the \emph{Fruit} dataset \cite{Heide}, which contains ten images of fruit, using Convolutional BCD (Algorithm \ref{algo:CBCD}), and use it to approximate the 11 test images (taken from \cite{LocalProcessing}), with local contrast normalization. Fig.~\ref{fig_psnr_types}(a) shows the PSNR as a function of the $\ell_{0,\infty}$ sparsity, averaged over the test images, for dictionaries trained for different target sparsities. It can be seen that the performance of the dictionaries improves significantly when they are trained for higher values of the $\ell_{0,\infty}$ ``norm''.

\begin{figure}[!t]
\captionsetup[subfigure]{justification=centering,font=scriptsize}
    \centering
  \subfloat[ ]{%
       \includegraphics[width=0.60\linewidth]{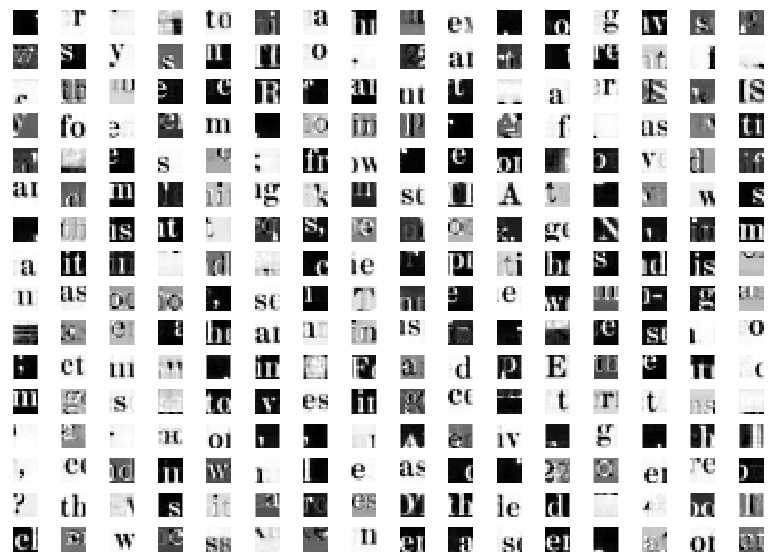}}
    \label{fig16a} \hfill
  \subfloat[ ]{ 
        \includegraphics[width=0.35\linewidth]{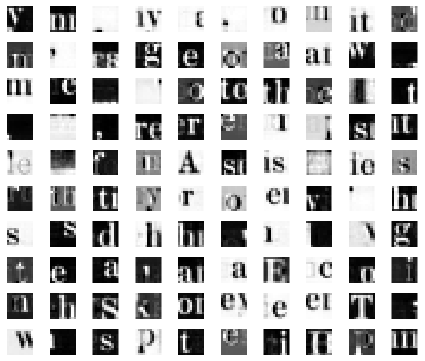}}
    \label{fig16b} 
  \caption{(a) A dictionary of 256 atoms and (b) a dictionary of 100 atoms, both trained on the clean training set.}
  \label{fig16}
\end{figure}

We compare the behavior of the dictionaries trained on natural images to the case of text images, which are of a different nature. Text images are very sparse in the $\ell_{0,\infty}$ sense, since such images, especially in typed documents, are composed of characters appearing at different shifts within the images. Thus, they can be approximated as a sum of convolutions between atoms representing individual characters and coefficients, which determine the locations of the characters within the image. 
In this case, each atom represents a whole character, and characters are positioned in non-overlapping locations. However, in most text layouts the spacing between characters, both horizontal and vertical, depends on the character. Therefore, a reasonable assumption would be that when representing a text image using atoms of single characters, there should be no more than one overlap between the atoms in any given pixel. This is equivalent to convolutional sparse coding with an $\ell_{0,\infty}$ ``norm'' equal to 2.

We used a dictionary of 100 atoms with size $11 \times 11$ (like the typical size of a character) trained on the 16 training images to represent the test set of 16 text images with various level of sparsity. Fig.~\ref{fig_textdict} shows the approximations of one of the images for different sparsity levels. We do not present histogram similar to Fig.~\ref{fig_hill_orig_cn} because the text dictionary, unlike the DCT atoms, cannot be sorted in a meaningful order of spatial frequencies. It can be observed that when the $\ell_{0,\infty}$ ``norm'' equals $2$, the image already looks good.
Fig.~\ref{fig_psnr_types}(b) shows the PSNR as a function of $\left\Vert \boldsymbol{\alpha} \right\Vert_{0,\infty}$, averaged over the test images, for dictionaries trained for different target sparsities. As expected, almost the same reconstruction accuracy is maintained on the test set even when the target sparsity is as low as two.
This is different from the natural images case, where we see a significant degradation when the dictionary target $\ell_{0,\infty}$ ``norm'' decreases.

Fig.~\ref{fig16} presents the atoms of the dictionary above (of size $100$) and compares it to a dictionary trained with 256 atoms. Both use a target $\ell_{0,\infty}$ ``norm'' of 2.
As expected, most atoms resemble individual characters and only few resemble a pair of them. 
As the $\ell_2$ cost function is used in updating the atoms, they do not uniformly represent the alphabet. Characters that appear frequently may have more than one atom resembling them, and characters that are rare may have no atom dedicated to them.  Note that the larger dictionary has more diversity.

\begin{figure}[!t]
\captionsetup[subfigure]{justification=centering,font=scriptsize}
\centering
  \subfloat[ ]{%
       \includegraphics[width=0.49\linewidth]{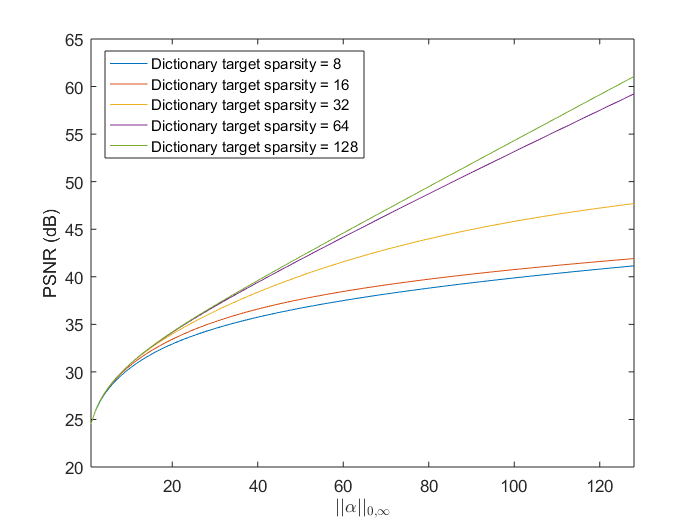}}
    \label{fig_psnr_types_a}
  \subfloat[ ]{ 
        \includegraphics[width=0.49\linewidth]{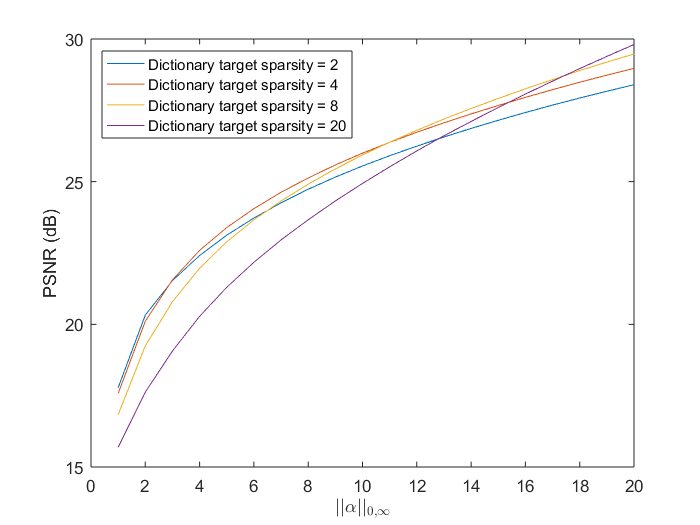}}
    \label{fig_psnr_types_b} 
\caption{PSNR as a function of $\left\Vert \boldsymbol{\alpha} \right\Vert_{0,\infty}$ using dictionaries trained for several target sparsities. The two image types are: (a) natural images with local contrast normalization; and (b) text images.}
\label{fig_psnr_types}
\end{figure}

\subsection{Convolutional sparse coding methods comparison}
We now compare the various greedy methods proposed. We used GCMP, GCOMP (Algorithm \ref{algo:GCOMP}), GCT (Algorithm \ref{algo:GCT}), and StGCOMP (Algorithm \ref{algo:StGCOMP} with $\Delta k = 2,4$) to represent the \emph{cameraman} image with $1 \leq \left\Vert \boldsymbol{\alpha} \right\Vert_{0,\infty} \leq 20$. Similar results are observed for other images. For comparison, we also used $\ell_0$ based convolutional matching pursuit \cite{CMP4}, \cite{CMP3}, and computed the $\ell_{0,\infty}$ norm at each iteration. 

\begin{figure}[!t]
\centering
\includegraphics[width=5.0in]{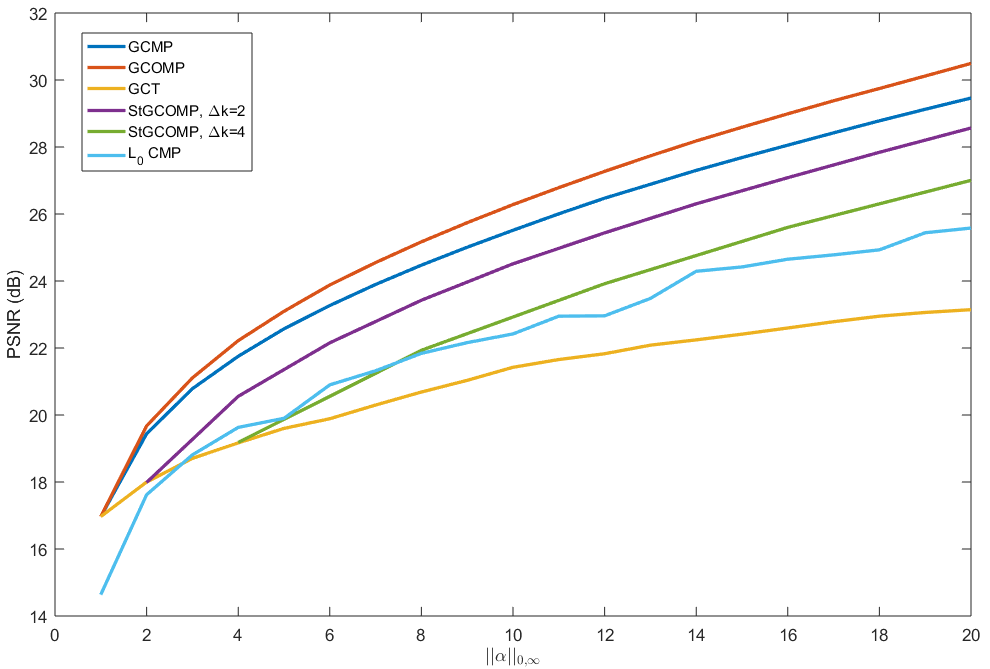}
\caption{PSNR as a function of $\left\Vert \boldsymbol{\alpha} \right\Vert_{0,\infty}$ for the proposed sparse coding methods.}
\label{fig2}
\end{figure}

Fig.~\ref{fig2} shows the PSNR as a function of the $\ell_{0,\infty}$ sparsity for the different algorithms. The PSNR was highest for GCOMP. Also, the PSNR of GCMP was lower than that of GCOMP because it does not include the additional step of optimizing the values of the coefficients after each iteration of the outer loop. GCT gave the worst reconstructions, as it greedily selects all the atoms at once. Also, Stagewise GCOMP gave PSNR values lower than GCOMP and higher than GCT, and $\Delta k = 2$ gave larger PSNR values than $\Delta k = 4$.

In the supplementary material, we measure the runtimes of GCMP and GCOMP (the two best algorithms according to this experiment), and compare them to the runtime of $\ell_0$ based convolutional matching pursuit \cite{CMP4}, \cite{CMP3}. In the sequel we use GCMP, as it offers the best trade-off between accuracy and runtime.

\subsection{Convolutional MOD vs. Convolutional BCD}
We trained a dictionary of 64 atoms of size $8 \times 8$ on the \emph{Fruit} dataset \cite{Heide}, which contains ten images of fruit, using Convolutional MOD (Algorithm \ref{algo:CMOD}) and using Convolutional BCD (Algorithm \ref{algo:CBCD}). Both dictionaries were constrained to $\left\Vert \boldsymbol{\alpha} \right\Vert_{0,\infty}=4$ and initialized to random Gaussian values. We computed the total squared error of the representations in each iteration. In both algorithms, the convergence rate is affected by the stopping condition of the CGLS dictionary update step, which we set to some mean squared error threshold $\epsilon$. For large values of $\epsilon$, each dictionary update step takes less time to compute, but achieves a smaller reduction in the squared error and therefore requires more iterations.

\begin{figure}[!t]
\centering
\includegraphics[width=5.0in]{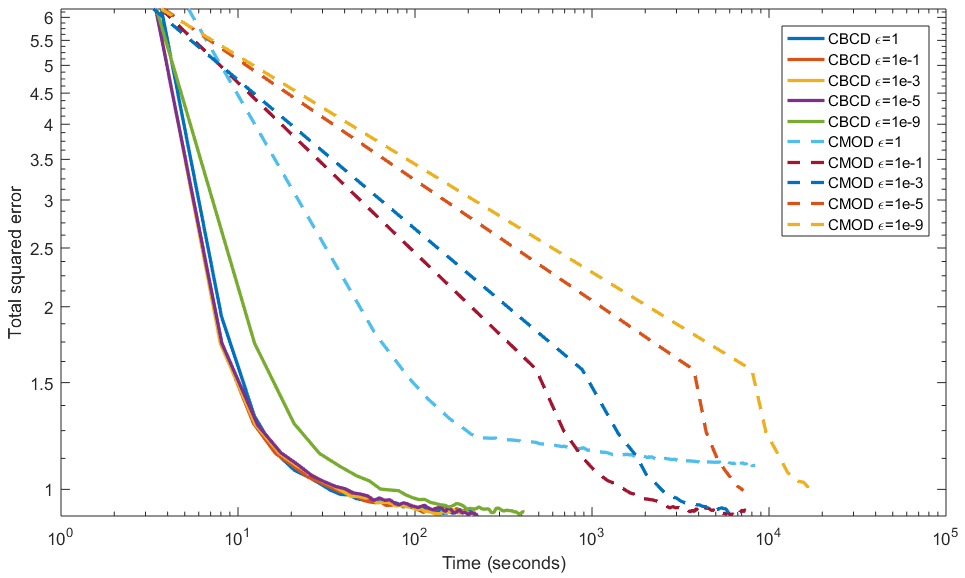}
\caption{Total squared error as a function of time for the two proposed dictionary learning algorithms, \emph{Fruit} dataset.}
\label{fig4}
\end{figure}

Fig.~\ref{fig4} compares the total squared error of the representations as a function of time for $\epsilon = 1, 10^{-1}, 10^{-3}, 10^{-5},$ and $10^{-9}$. The computation time of the first iteration of CMOD increased significantly with $\epsilon$, and resulted in a similar error reduction. In later iterations, CMOD converged slower for $\epsilon=1$ than for smaller values of $\epsilon$. Yet, the squared error of CBCD converged much faster than that of CMOD for all $\epsilon$ values and was less dependent on the value of $\epsilon$. 
Therefore, in the subsequent experiments we use CBCD with $\epsilon=1$.

In the supplementary material, we present visualizations of a dictionary trained for solving an $\ell_{0,\infty}$ constrained problem using Convolutional BCD, a dictionary trained for solving the standard $\ell_1$ convolutional sparse coding problem, and a dictionary trained on patches for solving an $\ell_0$ constrained problem using K-SVD.

\subsection{Inpainting}
\label{inpainting}
We next apply our algorithm on the inpainting problem of text data and natural images with the same setting presented in \cite{Heide,LocalProcessing}. We corrupt the images
by removing 50\% of the pixels from each image at random.
We use the pretrained and image specific inpainting methods from Section \ref{sectionInpainting} with a $\ell_{0,\infty}$ target sparsity of $64$.

We compare our results to the method from \cite{LocalProcessing} (using their software), which solves a $\ell_{0,\infty}$ problem with convex relaxation; and to the method from \cite{Heide} (using their software), which uses standard $\ell_1$ convolutional sparse coding. The PSNR (dB)  in this experiment is computed the same way as in \cite{LocalProcessing}, as $\text{PSNR}=20 \log\left(\frac{\sqrt{MN}}{\left\Vert \textbf{x}-\hat{\textbf{x}} \right\Vert_{2}}\right)$. We also apply our method for image specific inpainting, and compare our results to the image specific method from \cite{LocalProcessing} (using their software), which uses convolutional dictionary learning based on $\ell_{0,\infty}$ with convex relaxation; to the image specific method from \cite{Heide} (which is based on standard $\ell_1$ convolutional sparse coding) using the SPORCO library \cite{wohlberg-2016-sporco}, and to the patch-based method from \cite[Chapter 15]{MBook} (using their software), which uses K-SVD to train a dictionary on overlapping patches of the corrupted image. All techniques use dictionaries of the same size ($100$ atoms), are trained on the same data (see Section~\ref{sec:acc_vs_sparsity}), and process the same images with 50\% missing pixels.

Table \ref{table_inpainting} presents the PSNR of the inpainted text images, averaged over all the test images. Observe that 
our method leads to better results compared to the other strategies. We believe that this is due to the small target $\ell_{0,\infty}$ ``norm'' required for training a dictionary for text images.

Table \ref{table1} presents the results for the local contrast normalized natural images. For most images, our strategy has poorer results compared to those of \cite{LocalProcessing}, \cite{Heide} and \cite[Chapter 15]{MBook}. The methods in \cite{LocalProcessing} and \cite{Heide} use convex $\ell_1$ relaxations, which appear to be more successful at inpainting natural images than our $\ell_{0,\infty}$ method. Following the comparison in Section~\ref{sec:acc_vs_sparsity}, we believe that the reason for that is the higher $\ell_{0,\infty}$ norm required in the training of our dictionary in this case. Though the final target sparsity used in the sparse coding after the dictionary has been trained is the same in both text and natural images cases, the influence of the sparsity of the trained dictionary is strongly apparent.

\renewcommand{\arraystretch}{1.25}
\begin{table}
\centering
\scriptsize
\caption{Average PSNR (dB) of inpainted text images}
$
\begin{array}{|c|c|c|} \hline
$Method$ & $Pretrained dictionary$ & $Dictionary learning$ \\ \hline
\text{\cite[Ch. 15]{MBook}}  & 12.08 & 19.34 \\ 
\cite{Heide} & 19.19 & \textbf{23.30} \\
\cite{LocalProcessing} & 19.29 & 20.57  \\
$Our method$ & \textbf{19.78} & 21.88  \\ \hline
\end{array}
$
\label{table_inpainting}
\end{table}

\renewcommand{\arraystretch}{1.5}
\begin{table*}
\centering
\tiny
\caption{PSNR (dB) of inpainted images}
$
\begin{array}{|c|c|c|c|c|c|c|c|c|c|c|c|} \hline
\text{Method} & \text{barbara} & \text{boat} & \text{c.man} & \text{couple} & \text{finger} & \text{hill} & \text{house} & \text{lena} & \text{man} & \text{montage} & \text{peppers} \\ \hhline{|=|=|=|=|=|=|=|=|=|=|=|=|}
\text{\cite{Heide}, pretrained}  & 11.00 & 10.29 & 9.74 & 11.99 & 15.55 & 10.37 & 10.18 & 11.77 & 10.60 & 15.11 & 9.41 \\
\text{\cite{LocalProcessing}, pretrained} & \textbf{11.67} & \textbf{10.33} & \textbf{9.95} & \textbf{12.25} & \textbf{16.04} & \textbf{10.66} & \textbf{10.56} & \textbf{11.92} & \textbf{11.84} & \textbf{15.40} & 9.18 \\
\text{Our method, pretrained} & 10.92 & 9.89 & 9.57 & 10.71 & 13.88 & 9.67 & 9.60 & 11.33 & 10.62 & 13.72 & \textbf{9.61} \\
\hhline{|=|=|=|=|=|=|=|=|=|=|=|=|}
\text{\cite[Ch. 15]{MBook}, image specific} & 13.33 & 11.05 & 10.06 & 12.07 & 14.77 & 10.43 & 10.61 & 11.96 & 11.17 & 14.92 & 10.78 \\
\text{\cite{Heide}, image specific}  & \textbf{18.47} & \textbf{13.89} & \textbf{12.10} & \textbf{14.59} & 13.88 & \textbf{12.96} & \textbf{13.63} & \textbf{14.23} & \textbf{13.92} & \textbf{16.92} & \textbf{13.16} \\
\text{\cite{LocalProcessing}, image specific} & 15.20 & 11.60 & 10.68 & 12.41 & \textbf{16.07} & 10.90 & 11.77 & 12.35 & 11.71 & 15.67 & 11.45 \\
\text{Our method, image specific} & 13.18 & 10.94 & 9.63 & 11.59 & 15.44 & 10.25 & 10.91 & 11.43 & 10.87 & 14.38 & 10.44 \\
\hline
\end{array}
$
\label{table1}
\end{table*}

\subsection{Salt-and-pepper noise removal for text images}

Having seen the advantage of our method for inpainting of text images, we turn to demonstrate that this is the case also for another image processing application, namely, salt-and-paper noise removal.
  
Due to the type of the noise, when training on noisy images, some of the learned atoms in the dictionary trained on noisy images will contain only this noise. Thus, they can be sparsely represented using the impulse noise atom $\textbf{d}_0$. Therefore, we remove them automatically by solving the following minimization problem:
\begin{equation}
\underset{\boldsymbol{\alpha}}{\text{min}} \left\Vert \boldsymbol{\alpha} \right\Vert_0 ~s.t.~\left\Vert \textbf{d}_j - \textbf{d}_0 \ast \boldsymbol{\alpha} \right\Vert_2^2 \leq \epsilon,
\end{equation}
which indicates whether the atom is noise or not. 
If the $\ell_0$ ``norm'' of the atom $\textbf{d}_j$ is lower than a given threshold (which we set to 3), we prune it from the dictionary.

\begin{figure*}[!t]
\captionsetup[subfigure]{justification=centering,font=scriptsize}
    \centering
  \subfloat[ ]{
       \includegraphics[width=0.22\linewidth]{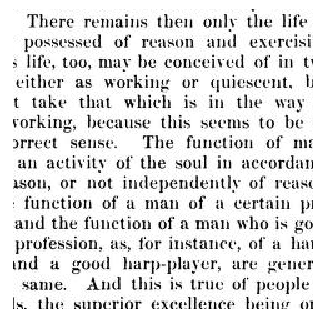}}
    \label{fig18a} \hfill
  \subfloat[ ]{ 
        \includegraphics[width=0.22\linewidth]{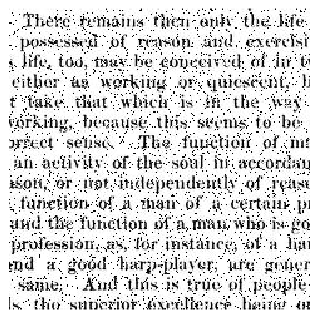}}
    \label{fig18b} \hfill
      \subfloat[ ]{ 
        \includegraphics[width=0.22\linewidth]{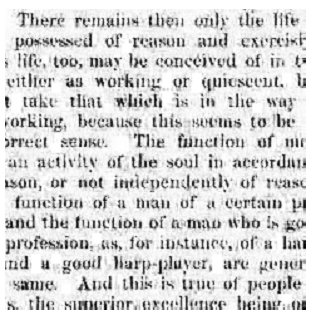}}
    \label{fig18c} \hfill
      \subfloat[ ]{ 
        \includegraphics[width=0.22\linewidth]{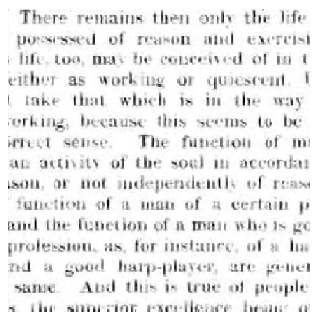}}
    \label{fig18d} \hfill
      \subfloat[ ]{ 
        \includegraphics[width=0.22\linewidth]{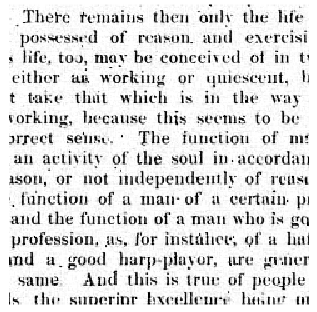}}
    \label{fig18e} \hfill
      \subfloat[ ]{ 
        \includegraphics[width=0.22\linewidth]{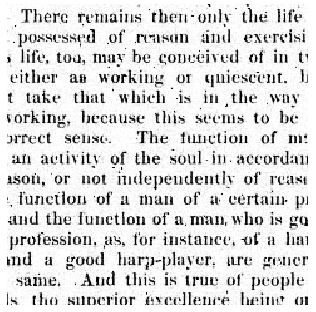}}
    \label{fig18f} \hfill
      \subfloat[ ]{ 
        \includegraphics[width=0.22\linewidth]{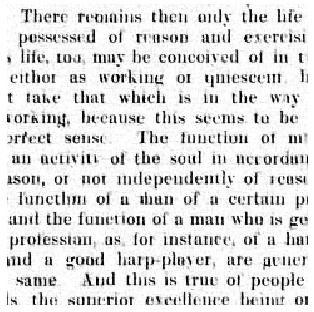}}
    \label{fig18g} \hfill
      \subfloat[ ]{ 
        \includegraphics[width=0.22\linewidth]{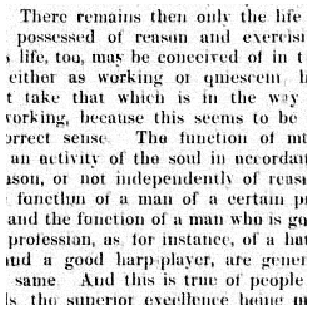}}
    \label{fig18h}
  \caption{Salt and pepper noise removal results for one of the test images: (a) the original (clean) image; (b) the noisy image; (c) denoising using the patch-based method from \cite{WeightedCouple}; (d) denoising using the convolutional sparse coding method from \cite{ImpulseNoise} (e) denoising using a dictionary trained on the clean training set, 100 atoms and (f) 256 atoms; (g) denoising using a dictionary trained on the noisy test set, 51 atoms and (h) 100 atoms.}
  \label{fig18}
\end{figure*}

We corrupted 10\% of the pixels of each test image by setting 5\% of the pixels to black and 5\% of the pixels to white at random. Fig.~\ref{fig18}(b) shows an example of a noisy image and Fig.~\ref{fig18}(a) shows its original clean version. We applied Algorithm \ref{algo:SNP}, both in the case of training on a clean training set and training on noisy images. 
As a reference, we compare our results to (i) a simple $3 \times 3$ median filter; (ii) the \emph{weighted couple sparse representation} method from \cite{WeightedCouple} (using their software), which trains a dictionary on overlapping patches by solving a weighted rank-one minimization problem; and (iii) the method from \cite{ImpulseNoise} (using their software), which is based on standard $\ell_1$ convolutional sparse coding. All methods other than the median filter trained dictionaries with 100 atoms of size $11 \times 11$ on the same noisy text images. We have tuned the relevant hyperparameters of each method to give the best results for the test images.

Table \ref{table3} presents the average PSNR of the different denoising methods and the initial average PSNR of the noisy images. The best results are achieved using our method with the (larger) dictionary of 100 atoms that was trained on the noisy test set. 
Fig.~\ref{fig18} presents the results for one of the test images. Simple methods such as a median filter do not work well in the case of text images due to considerable degradation of the clean parts of the images (and larger filter sizes cause more degradation). The patch-based method from \cite{WeightedCouple} and the convolutional method from \cite{ImpulseNoise}, which work well for natural images, also degrade the clean parts of the text images, while our method provides better results.

\renewcommand{\arraystretch}{1.25}
\begin{table}
\centering
\scriptsize
\caption{Average PSNR of denoised images, 10\% noise}
$
\begin{array}{|c|c|} \hline
\text{Method} & \text{PSNR (dB)} \\ \hline
\text{Noisy image} & 13.37 \\
\text{Median filter} & 16.53 \\
\text{Patch-based method \cite{WeightedCouple}} & 13.27 \\
\text{Convolutional sparse coding method from \cite{ImpulseNoise}} & 16.63 \\
\text{Train on clean training set, 100 atoms} & 21.60 \\
\text{Train on clean training set, 256 atoms} & 21.43 \\
\text{Train on noisy test set, 51 atoms} & 21.36 \\
\text{Train on noisy test set, 100 atoms} & 22.30 \\ \hline
\end{array}
$
\label{table3}
\end{table}

\section{Conclusions}
\label{sec:conclusions}
The greedy algorithms that we have proposed in this work solve the constrained $P_{0,\infty}^\epsilon$ problem and offer an alternative to the approaches in \cite{LocalProcessing} and \cite{WorkingLocally}, which minimize an unconstrained penalized Lagrangian with a convex relaxation to the $\ell_1$ norm. One advantage of using greedy algorithms over relaxation based techniques is that they allow minimizing the squared error with a hard constraint on the $\ell_{0,\infty}$ sparsity or minimizing the $\ell_{0,\infty}$ sparsity with a hard constraint on the squared error. Our pursuits are accompanied by a dictionary learning method that uses our greedy solution to $P_{0,\infty}^\epsilon$ as the sparse coding step separately from the dictionary update step. This allows us to train dictionaries on a set of training signals with a target sparsity. Our techniques are computationally efficient and very easy to implement.

Since our greedy approach targets solely the $\ell_{0,\infty}$ ``norm'', and not in conjunction with other norms as in \cite{LocalProcessing}, we believe that it serves as a good tool to evaluate this `'norm'' for various image types. From our experiments, it seems that while targeting this norm directly is beneficial for text images, which are very sparse in this ``norm'', this is less the case for natural images. For the latter, it seems that it may be more beneficial to use the $\ell_{0,\infty}$ sparsity in conjunction with other priors such as the $\ell_1$ norm, and not alone. Clearly, this conclusion may be specific to the greedy optimization technique we have proposed.

\section*{Acknowledgments}
The authors are grateful to the reviewers' valuable comments that improved the manuscript.
This work was supported by ERC-StG grant no. 757497 (SPADE).

\bibliographystyle{siamplain}
\bibliography{refs}
\end{document}

%% file: rev_ex_shared.tex
\usepackage{amsfonts}
\usepackage{graphicx}
\usepackage{epstopdf}
\usepackage{mathtools}
\usepackage{amsmath}
\usepackage{amssymb}
\usepackage{algorithm}
\usepackage{algpseudocode}
\usepackage{array}
\usepackage{comment}
\usepackage{hhline}
\usepackage[caption=false,font=footnotesize]{subfig}
\usepackage{tikz}
\usetikzlibrary{spy}
\ifpdf
  \DeclareGraphicsExtensions{.eps,.pdf,.png,.jpg}
\else
  \DeclareGraphicsExtensions{.eps}
\fi

\numberwithin{theorem}{section}

\newcommand{\TheTitle}{A Greedy Approach to $\ell_{0,\infty}$ Based Convolutional Sparse Coding} 
\newcommand{\TheAuthors}{E. Plaut and R. Giryes}

\headers{\TheTitle}{\TheAuthors}

\title{{\TheTitle}\thanks{Accepted for publication in SIAM Journal on Imaging Sciences (SIIMS).}}

\author{
  Elad Plaut\thanks{School of Electrical Engineering, Tel Aviv University, Tel Aviv,
Israel   (\email{eladplaut@mail.tau.ac.il}).}
  \and
  Raja Giryes\thanks{School of Electrical Engineering, Tel Aviv University, Tel Aviv,
Israel (\email{raja@tauex.tau.ac.il}).}
}

\usepackage{amsopn}
